\documentclass[twocolumn,preprintnumbers,amsmath,amssymb,showpacs]{revtex4}
\usepackage{epsfig}
\usepackage{color}
\begin{document}
\setlength{\voffset}{1.0cm}
\title{Large $N$ solution of generalized Gross-Neveu model with two coupling constants}
\author{Christian Boehmer\footnote{christian.boehmer@theorie3.physik.uni-erlangen.de}}
\author{Michael Thies\footnote{thies@theorie3.physik.uni-erlangen.de}}
\affiliation{Institut f\"ur Theoretische Physik III,
Universit\"at Erlangen-N\"urnberg, D-91058 Erlangen, Germany}
\date{\today}
\begin{abstract}
The Gross-Neveu model in 1+1 dimensions is generalized to the case of different scalar and pseudoscalar coupling constants.
This enables us to interpolate smoothly between the standard massless Gross-Neveu models with either discrete or continuous chiral symmetry.
We present the solution of the generalized model in the large $N$ limit including the vacuum, fermion-antifermion scattering and bound states,
solitonic baryons with fractional baryon number and the full phase diagram at finite temperature and chemical potential. 
\end{abstract}
\maketitle

\section{Introduction}\label{sect1}
The sustained interest in Gross-Neveu (GN) models in 1+1 dimensions \cite{1} stems to a large extent from their chiral properties. Thus
the simplest model with Lagrangian    
\begin{equation}
{\cal L}=\bar{\psi} {\rm i}\gamma^{\mu}\partial_{\mu}\psi + \frac{1}{2} g^2 (\bar{\psi}\psi)^2  
\label{1.1}
\end{equation}
(suppressing flavor indices, i.e., $\bar{\psi}\psi = \sum_{k=1}^N \bar{\psi}_k\psi_k$ etc.) has a discrete chiral Z$_2$ symmetry
\begin{equation}
\psi \to \gamma_5 \psi,
\label{1.2}
\end{equation}
whereas the chiral GN model or, equivalently, the two-dimensional Nambu--Jona-Lasinio model (NJL$_2$) \cite{2},
\begin{equation}
{\cal L}=\bar{\psi} {\rm i}\gamma^{\mu}\partial_{\mu}\psi + \frac{1}{2} g^2 (\bar{\psi}\psi)^2 + \frac{1}{2} g^2 (\bar{\psi}{\rm i}\gamma_5 \psi)^2,
\label{1.3}
\end{equation}
possesses a continuous chiral U(1) symmetry,
\begin{equation}
\psi \to {\rm e}^{{\rm i}\alpha \gamma_5} \psi.
\label{1.4}
\end{equation}
Chiral symmetry and in particular its breakdown manifest themselves in such diverse physical phenomena as dynamical fermion masses, the meson
spectrum, topological effects in the structure of baryons, and rich phase diagrams at finite density and temperature with various types of homogeneous and
solitonic crystal 
phases, see the introductory review article \cite{3} as well as the recent updates in \cite{4,5,6}.  
By adding a bare mass term to the Lagrangian, one breaks the chiral symmetry explicitly and gets additional insights into the symmetry aspects of both 
models \cite{7,8}. Nevertheless, studies of models (\ref{1.1}) and (\ref{1.3}) with their strikingly different properties have remained somewhat disconnected.

In the present work, we propose and solve a simple field theoretical model which interpolates continuously between the Lagrangians (\ref{1.1}) and (\ref{1.3}).
Our motivation is to get a better understanding of how the conspicuous differences in the phase diagrams and baryon structure come about. Moreover,
we would like to explore an alternative mechanism for breaking chiral symmetry explicitly, different from the usual bare mass term. To this end, we
consider a Lagrangian similar to Eq.~(\ref{1.3}), but with different (attractive) scalar and pseudoscalar couplings,
\begin{equation}
{\cal L}=\bar{\psi} {\rm i}\gamma^{\mu}\partial_{\mu}\psi + \frac{1}{2} g^2 (\bar{\psi}\psi)^2 + \frac{1}{2} G^2 (\bar{\psi}{\rm i}\gamma_5 \psi)^2.
\label{1.5}
\end{equation}
By varying $G^2$ from 0 to $g^2$, we generate a family of theories interpolating between the GN and the NJL$_2$ models. The idea to 
generalize the GN model in this fashion is not new. Thus for instance, Klimenko has studied a closely related problem long time ago \cite{9,10}.
However, since the
role of inhomogeneous condensates has only been appreciated in recent years, there is almost no overlap between the present work
and these earlier studies.

The methods which we shall use in our investigation have been developed during the last few years in an effort to clarify the phase structure
of massless and massive GN models. As a result, we have now at our disposal a whole toolbox of analytical and numerical instruments.
The most important keywords are: the derivative expansion, asymptotic expansions, perturbation theory, Ginzburg-Landau (GL)
theory and numerical Hartree-Fock (HF) approach including the Dirac sea. This will enable us   
to solve the generalized GN model (\ref{1.5}) in a rather straightforward fashion, although the model is far from trivial. Its two limiting cases,
the standard massless GN and NJL$_2$ models, can both be solved analytically. This is unfortunately not true for the generalized model
which in this respect is closer to the massive NJL$_2$ model \cite{8}.

This paper is organized as follows. We present our computations and results starting with mostly analytical work and ending with purely 
numerical results. The logic of the HF approach demands that we begin with a discussion of  the vacuum, dynamical fermion mass and coupling
constant renormalization in Sec.~\ref{sect2}. Sec.~\ref{sect3} is dedicated to fermion-fermion bound states (mesons) and scattering. In Sec.~\ref{sect4},
we solve the theory in the baryon sector as well as for low density soliton crystals in the vicinity of the chiral limit, using a kind of chiral perturbation
theory obtained from the derivative expansion. We then begin our study of thermodynamics at finite temperature and chemical potential with an
investigation of the tricritical behavior near the chiral limit in Sec.~\ref{sect5}. In Sec.~\ref{sect6} the microscopic GL approach underlying Sec.~\ref{sect5}
is extended to more general coupling constants, and the tricritical point of the generalized GN model is determined exactly.
Some technical details are deferred to the appendix.
Sec.~\ref{sect7} is devoted to the full phase diagram of the generalized GN model for arbitrary coupling constants, chemical potential and temperature, 
only accessible via a numerical relativistic HF calculation. As a by-product, we also present information about baryons away from the chiral limit. 
The paper ends with a concluding section, Sec.~\ref{sect8}. 
 
\section{Vacuum, dynamical fermion mass, renormalization}\label{sect2}

Consider the Lagrangian of the generalized GN model with two coupling constants in 1+1 dimensions, Eq.~(\ref{1.5}).
For $G^2=g^2$, it coincides
with the one from the massless NJL$_2$ model, Eq.~(\ref{1.3}). For $G^2=0$, we recover
the massless GN model, Eq.~(\ref{1.1}). The case $G^2>g^2$ can be mapped 
onto $G^2<g^2$ by means of a chiral rotation about a quarter of a circle,
\begin{equation}
\psi \to {\rm e}^{{\rm i}\gamma_5 \pi/4}\psi.
\label{2.1}
\end{equation}
Since this is a canonical transformation, we may assume $0<G^2<g^2$ without loss of generality.
Hence the generalized GN model can serve as a continuous interpolation between two well-studied model field theories with distinct symmetry properties.
Notice that the generalized Lagrangian (\ref{1.5}) always has the discrete chiral symmetry $\psi \to \gamma_5 \psi$ under which $\bar{\psi}\psi$
and $\bar{\psi}{\rm i}\gamma_5 \psi$ change sign. The continuous chiral symmetry $\psi \to {\rm e}^{{\rm i} \alpha \gamma_5} \psi$ is only recovered 
at the point $g^2=G^2$. 

To find the vacuum in the large $N$ limit, we introduce homogeneous scalar and pseudoscalar condensates,
\begin{eqnarray}
m & = & - g^2 \langle \bar{\psi}\psi \rangle,
\nonumber \\
M & = & - G^2 \langle \bar{\psi}{\rm i} \gamma_5 \psi \rangle.
\label{2.2}
\end{eqnarray}
The Dirac-Hartree-Fock equation 
\begin{equation}
\left( - \gamma_5 {\rm i} \partial_x + \gamma^0 m + {\rm i} \gamma^1 M \right)\psi = E \psi
\label{2.3}
\end{equation}
then yields the single particle energies
\begin{equation}
E = \pm \sqrt{k^2+m^2+M^2}
\label{2.4}
\end{equation}
and the (cutoff regularized) vacuum energy,
\begin{eqnarray}
{\cal E}_{\rm vac} & = & - \int_{-\Lambda/2}^{\Lambda/2} \frac{{\rm d}k}{2\pi} \sqrt{k^2+m^2+M^2} + \frac{m^2}{2Ng^2} +\frac{M^2}{2NG^2}
\nonumber \\
& = & - \frac{\Lambda^2}{8\pi} + \frac{m^2+M^2}{4\pi}\left[ \ln \left(\frac{m^2+M^2}{\Lambda^2}\right)-1 \right] 
\nonumber \\
& & + \frac{m^2}{2Ng^2} +\frac{M^2}{2NG^2}.
\label{2.5}
\end{eqnarray}
If we choose the following relations between the UV cutoff $\Lambda/2$ and the bare coupling constants $g^2,G^2$,
\begin{eqnarray}
\frac{\pi}{Ng^2} - \ln \Lambda & = &  \xi_1 ,
\nonumber \\
\frac{\pi}{NG^2} - \ln \Lambda & = &  \xi_2 , 
\label{2.6}
\end{eqnarray}
${\cal E}_{\rm vac}(m,M)$ is well defined in the limit $\Lambda \to \infty$ (dropping the irrelevant quadratic divergence) and given by
\begin{equation}
{\cal E}_{\rm vac} = \frac{m^2+M^2}{4\pi} \left[ \ln \left(m^2+M^2 \right)-1 \right] + \frac{\xi_1 m^2}{2\pi} + \frac{\xi_2 M^2}{2\pi}. 
\label{2.7}
\end{equation}
Minimize ${\cal E}_{\rm vac}$ with respect to $m,M$,
\begin{eqnarray}
0 & = & m \left[ 2 \xi_1 + \ln \left( m^2+M^2 \right) \right],
\nonumber \\
0 & = & M \left[ 2 \xi_2 + \ln \left( m^2+M^2 \right) \right].
\label{2.8}
\end{eqnarray}
These equations only admit a solution with nonvanishing $m$ and $M$ if $\xi_1=\xi_2=- \frac{1}{2}\ln(m^2+M^2)$. This takes us back to the
NJL$_2$ model with its infinitely degenerate vacua along the chiral circle of radius $\sqrt{m^2+M^2}$. The other options are
$m\neq 0, M=0, \xi_2$ unspecified and
\begin{equation}
\xi_1 = - \frac{1}{2} \ln m^2, \qquad {\cal E}_{\rm vac} = - \frac{m^2}{4\pi},
\label{2.9}
\end{equation}
or else $m=0, M\neq 0, \xi_1$ unspecified and
\begin{equation}
\xi_2 = -  \frac{1}{2} \ln M^2, \qquad {\cal E}_{\rm vac} = - \frac{M^2}{4\pi}.
\label{2.20}
\end{equation}
The vacuum energy is lowest for $m \neq 0$ if $\xi_1<\xi_2$ and for $M \neq 0$ if $\xi_1>\xi_2$. In view of the remark below Eq.~(\ref{2.1}), we may adopt
the first scenario. Choosing units such that $m=1$ and denoting $\xi_2 (>0)$ by $\xi$ from now on, we finally get the
renormalization conditions (gap equations)
\begin{eqnarray}
\frac{\pi}{Ng^2} &=&  \ln \Lambda,
\nonumber \\
\frac{\pi}{NG^2} & = & \xi + \frac{\pi}{Ng^2}  = \xi + \ln \Lambda.
\label{2.11}
\end{eqnarray}
With the help of these relations, all physical quantities can be expressed in terms of the scale $m$ (set equal to 1) and the dimensionless parameter $\xi$ 
which serves to interpolate between the massless NJL$_2$ ($\xi=0$) and GN ($\xi=\infty$) models. 
This expectation is borne out in the following sections, supporting our renormalization method.

\section{Meson spectrum and fermion-antifermion scattering}\label{sect3}

\begin{figure}
\begin{center}
\epsfig{file=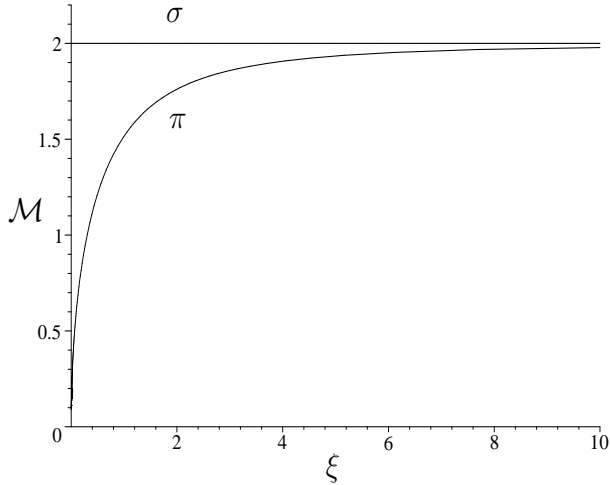,height=8cm,width=6.4cm,angle=270}
\caption{Masses of $\sigma$ and $\pi$ mesons vs. $\xi$ in the large $N$ limit of the generalized GN model, obtained from 
Eqs.~(\ref{3.7}-\ref{3.10}).}
\label{fig1}
\end{center}
\end{figure}

In the large $N$ limit, fermion-antifermion bound  and scattering states can conveniently be derived via the relativistic random phase
approximation (RPA) \cite{11,12}. Since the scalar and pseudoscalar channels decouple and the HF vacuum is the same as in the GN or NJL$_2$ model,
this analysis requires only minor changes of the standard calculation for the NJL$_2$ model. Consider first the bound state problem. 
The scalar channel has been spelled out in all detail in Ref.~\cite{12} where it is shown that the eigenvalue equation assumes the form
\begin{eqnarray}
1 &=& 2 Ng^2 \int \frac{{\rm d}k}{2\pi} \bar{u}(k)v(k-P) \bar{u}(k-P)v(k)
\nonumber \\
& &  \times \frac{E(k-P,k)}{{\cal E}^2(P)-E^2(k-P,k)}
\label{3.1}
\end{eqnarray}
Here, $P$ is the total momentum of the fermion-antifermion system, $u,v$ are positive and negative energy HF spinors, and 
\begin{equation}
E(k)=\sqrt{k^2+1}, \qquad E(k',k) = E(k')+E(k).
\label{3.2}
\end{equation}
The energy of the meson is denoted by ${\cal E}(P)= \sqrt{P^2 + {\cal M}^2}$. 
An analogous computation in the pseudoscalar channel gives 
\begin{eqnarray}
1 &=& - 2 NG^2 \int \frac{{\rm d}k}{2\pi} \bar{u}(k){\rm i}\gamma_5 v(k-P) \bar{u}(k-P){\rm i}\gamma_5 v(k)
\nonumber \\
& & \times  \frac{E(k-P,k)}{{\cal E}^2(P)-E^2(k-P,k)}.
\label{3.3}
\end{eqnarray}
Use of the identities
\begin{equation}
\bar{u}(k)v(k-P)\bar{u}(k-P)v(k)  =  \frac{4+P^2-E^2(k-P,k)}{4E(k)E(k-P)} 
\label{3.4}
\end{equation}
\begin{equation}
\bar{u}(k){\rm i}\gamma_5 v(k-P)\bar{u}(k-P){\rm i}\gamma_5 v(k)  = - \frac{P^2-E^2(k-P,k)}{4E(k)E(k-P)}   
\label{3.5}
\end{equation}
puts these eigenvalue equations into the more convenient form
\begin{eqnarray}
1&=&\frac{Ng^2}{2} \int \frac{{\rm d}k}{2\pi} \left( \frac{1}{E(k-P)}+\frac{1}{E(k)}\right)
\nonumber \\
& & \times \frac{4+P^2-E^2(k-P,k)}{{\cal E}^2(P)- E^2(k-P,k)},
\nonumber \\
1 & = & \frac{NG^2}{2} \int \frac{{\rm d}k}{2\pi} \left( \frac{1}{E(k-P)}+\frac{1}{E(k)}\right)
\nonumber \\
& & \times \frac{P^2-E^2(k-P,k)}{{\cal E}^2(P)- E^2(k-P,k)}.
\label{3.6}
\end{eqnarray}
If we regularize the momentum integrals with the same cutoff $\Lambda/2$ as used in the treatment of the vacuum energy and use the 
renormalization conditions Eqs.~(\ref{2.11}), we get the renormalized eigenvalue conditions
\begin{eqnarray}
0 & = & \int \frac{{\rm d}k}{2\pi} \left( \frac{1}{E(k-P)}+\frac{1}{E(k)}\right)
\nonumber \\
& & \times  \frac{4+P^2-{\cal E}^2(P)}{{\cal E}^2(P)- E^2(k-P,k)},
\label{3.7} \\
\frac{2\xi}{\pi} & = &  \int \frac{{\rm d}k}{2\pi} \left( \frac{1}{E(k-P)}+\frac{1}{E(k)}\right)
\nonumber \\
& & \times \frac{P^2-{\cal E}^2(P)}{{\cal E}^2(P)- E^2(k-P,k)},
\label{3.8}
\end{eqnarray}
now free of divergences.
Eq.~(\ref{3.7}) is the same as in the standard GN and NJL$_2$ models and gives the familiar result for the scalar ($\sigma$) meson mass, ${\cal M}=2$.
The right-hand side of Eq.~(\ref{3.8}) is independent of $P$ and can readily be evaluated in the cm frame of the meson ($P=0$),
\begin{eqnarray}
\xi &=&  - \frac{{\cal M}^2}{2} \int {\rm d}k \frac{1}{\sqrt{k^2+1}({\cal M}^2-4-4k^2)} 
\nonumber \\
& = &  \frac{1}{\sqrt{\eta-1}}\arctan \frac{1}{\sqrt{\eta-1}} 
\label{3.9}
\end{eqnarray}
with
\begin{equation}
\eta = \frac{4}{{\cal M}^2}
\label{3.10}
\end{equation}
Solving the transcendental equation (\ref{3.9}) numerically, the pseudoscalar ($\pi$) meson mass is found to rise from ${\cal M}=0$ at $\xi=0$ to 2 at $\xi\to \infty$,
see Fig.~\ref{fig1}. The first limit is as expected -- this is the would-be Goldstone boson of the NJL$_2$ model. The 2nd one is
surprising at first glance, since we are supposed to reach the GN model in this limit. The GN model does not have any pseudoscalar fermion-antifermion
interaction, let alone a bound state. 

To better understand what is going on, we briefly turn to the fermion-antifermion scattering problem. Since the RPA equations have a separable kernel
with one-term separable potentials in the scalar and pseudoscalar channels, this is straightforward \cite{13}. The energy dependence of the 
scattering matrix is encoded in the following functions of the Mandelstam variable $s$,
\begin{eqnarray}
\tau_{\sigma} & = & \frac{Ng^2}{1+ N g^2 \int \frac{{\rm d}k}{2\pi} \frac{1}{\sqrt{1+k^2}}\frac{4 k^2}{s-4(1+ k^2)+{\rm i}\epsilon}}
\nonumber \\
\tau_{\pi} & = & \frac{NG^2}{1+ N G^2 \int \frac{{\rm d}k}{2\pi} \frac{1}{\sqrt{1+k^2}}\frac{4(1+ k^2)}{s-4(1+ k^2)+{\rm i}\epsilon}}
\label{3.11}
\end{eqnarray}
Upon isolating the divergent part of the integrals and using the renormalization conditions, this becomes
\begin{eqnarray}
\tau_{\sigma}^{-1} & = & \frac{(s-4)}{2\pi} I(s)
\nonumber \\
\tau_{\pi}^{-1} & = & \frac{\xi}{\pi} + \frac{s}{2\pi} I(s)
\nonumber \\
I(s) & = & \int {\rm d}k \frac{1}{\sqrt{1+k^2}}\frac{1}{s-4(1+k^2)+{\rm i} \epsilon}
\label{3.12}
\end{eqnarray}
where the integral $I(s)$ can be evaluated in closed form,
\begin{equation}
I(s) =  - \frac{2}{\sqrt{s(4-s)}} \arctan \sqrt{\frac{s}{4-s}} \quad (s<4)
\label{3.13}
\end{equation}
\begin{equation}
I(s)  =  \frac{1}{\sqrt{s(s-4)}} \left( \ln \frac{\sqrt{s}+\sqrt{s-4}}{\sqrt{s}-\sqrt{s-4}} - {\rm i}\pi \right) \quad (s>4)
\label{3.14}
\end{equation}
$\tau_{\sigma}$ has the expected pole at $s=4$ corresponding to the marginally bound scalar meson with ${\cal M}=2$.
The pole of $\tau_{\pi}$ in turn coincides with the mass of the pseudoscalar meson, see Eqs.~(\ref{3.9},\ref{3.10}).
According to the 2nd line of Eq.~(\ref{3.12}), the strength of the pseudoscalar scattering matrix vanishes like $\sim 1/\xi$
for $\xi \to \infty$. We therefore arrive at the following picture: As $\xi \to \infty$, the pseudoscalar interaction vanishes,
in accordance with the expected GN limit. However, since an arbitrary weak attractive interaction is sufficient to support
a bound state in 1+1 dimensions, the pseudoscalar bound state pole persists, the binding energy going to zero.
As we shall see later on, this decoupled $\pi$ meson has no influence on any other observables of the model in the large
$N$ limit, so that it does not really upset our goal of interpolating between the NJL$_2$ and GN models.

\section{Baryons and soliton crystals at small $\xi$ and low density}\label{sect4}

The derivative expansion is a standard technique to deal with quantum mechanical particles subject to smooth potentials \cite{14,15}. In Ref.~\cite{16} it has
been adapted to the particular needs of the HF approach for low dimensional fermion field theories. In effect, it amounts to integrating out the
fermions in favor of an effective bosonic field theory, where the scalar and pseudoscalar fields can be identified with the 
HF potentials related to the composite fermion operators $\bar{\psi}\psi$ and $\bar{\psi}{\rm i}\gamma_5 \psi$. For baryons in the massive NJL$_2$ model
it leads to a chiral expansion in closed analytical form \cite{16}. Note that this method can only handle fully occupied valence levels at present.

Since the HF equation in the problem at hand has the same form as in the NJL$_2$ model, we can take over the derivation of the effective
action from Ref.~\cite{16} almost literally. The Dirac-HF equation is written as in Eqs.~(\ref{2.2},\ref{2.3}) except that the scalar
($S$) and pseudoscalar ($P$) condensates in the baryon state are $x$ dependent,
\begin{equation}
\left[ - \gamma_5 {\rm i} \partial_x + \gamma^0 S(x) + {\rm i} \gamma^1 P(x) \right] \psi = E \psi,
\label{2.3a}
\end{equation}
with
\begin{eqnarray}
S & = & - g^2 \langle \bar{\psi}\psi \rangle,
\nonumber \\
P & = & - G^2 \langle \bar{\psi}{\rm i} \gamma_5 \psi \rangle.
\label{2.2a}
\end{eqnarray}
As is well known, the HF energy can be written as the sum over single particle energies of occupied orbits 
and a double counting correction. Only
this last part is different in the present case. Due to the renormalization condition (\ref{2.11}), it depends on the parameter $\xi$,
\begin{equation}
{\cal E}_{\rm d.c.} = \frac{S^2}{2Ng^2} + \frac{P^2}{2NG^2} = \frac{S^2+P^2}{2\pi} \ln \Lambda + \frac{\xi}{2\pi} P^2.
\label{4.1}
\end{equation}
The cutoff dependent term cancels exactly the logarithmic divergence in the sum over single particle energies. Only the last term in Eq.~(\ref{4.1})
is different from what it was before. Consequently, we can
simply take over the effective action from Ref.~\cite{16}, set the confinement parameter $\gamma=0$ (vanishing bare fermion mass) and add the new contribution
proportional to $\xi$ from Eq.~(\ref{4.1}). Adopting polar coordinates in field space,
\begin{equation}
S-{\rm i}P = (1+\lambda) {\rm e}^{2{\rm i}\chi},
\label{4.2}
\end{equation}
and working at the same order in the derivative expansion as in \cite{16}, we then get at once the energy density ($' = \partial_x$ and $\chi^{IV}$
denotes the 4th derivative of $\chi$)
\begin{eqnarray}
2 \pi {\cal E} &=& \xi (1+\lambda)^2\sin^2 (2\chi)
 +  (\chi')^2 - \frac{1}{6} (\chi'')^2 + \frac{1}{30}(\chi''')^2
\nonumber \\
&- &  \frac{1}{140}(\chi^{IV})^2  - \frac{1}{45} (\chi'')^4 +  \lambda^2 + \frac{1}{12} (\lambda')^2  + \frac{1}{3} \lambda^3 
\nonumber \\
&- &
  \frac{1}{120}(\lambda'')^2 - \frac{1}{6} \lambda (\lambda')^2 - \frac{1}{12} \lambda^4 + \frac{1}{3} \lambda (\chi'')^2
\nonumber \\
&+ &   \frac{1}{15} \lambda (\chi''')^2 + \frac{1}{5}\lambda \chi'' \chi^{IV}
  - \frac{1}{2} \lambda^2 (\chi'')^2.
\label{4.3}
\end{eqnarray}
We have to vary the energy functional with respect to $\lambda$ and $\chi$ and solve the Euler-Lagrange equations, then 
compute baryon number and baryon mass. Although we shall follow the same procedure as in Ref.~\cite{16}, the results will be
quite different, reflecting the different ways in which chiral symmetry is broken in these two models.
For simplicity, take first the case of the leading order (LO) derivative expansion. Here, we only keep two terms in the energy
density,
\begin{equation}
2\pi {\cal E} = \xi \sin^2(2\chi) + (\chi')^2.
\label{4.4}
\end{equation}
Rescaling the chiral phase field and its spatial argument as follows,
\begin{equation}
\chi(x)=\frac{1}{4} \theta(y), \qquad y=2 \sqrt{\xi} x,
\label{4.5}
\end{equation}
we recognize the (static) sine-Gordon action ($\dot{\ }=\partial_y$)
\begin{equation}
\frac{4\pi}{\xi} {\cal E} = \frac{1}{2} \dot{\theta}^2 - \cos \theta +1.
\label{4.6}
\end{equation}
The Euler-Lagrange equation is the time-independent sine-Gordon equation
\begin{equation}
\ddot{\theta} = \sin \theta,
\label{4.7}
\end{equation}
so that the baryon can be identified with the sine-Gordon kink 
\begin{equation}
\theta= 4 \arctan {\rm e}^y.
\label{4.8}
\end{equation}
But unlike in the massive NJL$_2$ model, this object has baryon number 1/2, exactly like the kink in the standard GN model (with fully occupied 
zero-mode),
\begin{equation}
N_{\rm B} = \int {\rm d}x \frac{\chi'}{\pi} = \frac{1}{\pi} \left[\chi(\infty)- \chi(-\infty)\right] = \frac{1}{2}.
\label{4.9}
\end{equation}
Here we have used the topological relationship between baryon number and winding number of the chiral phase \cite{11,16}.
The mass of this kink-like baryon is found to be
\begin{equation}
\frac{M_{\rm B}}{N}= \frac{\sqrt{\xi}}{\pi} = \frac{m_{\pi}}{2\pi},
\label{4.10}
\end{equation}
where, in the 2nd step, we have made use of Eq.~(\ref{3.9}) to LO in $\xi$ and denoted the pion mass by $m_{\pi}$.

In the same vein, higher order calculations closely follow Ref.~\cite{16}. We find it useful to switch from the parameter $\xi$ to $m_{\pi}$
by means of Eq.~(\ref{3.9}),
\begin{equation}
\xi \approx \frac{1}{4} m_{\pi}^2 + \frac{1}{24} m_{\pi}^4 + \frac{1}{120} m_{\pi}^6 + \frac{1}{560} m_{\pi}^8,
\label{4.11}
\end{equation}
and to expand $\chi$ and $\lambda$ into Taylor series in $m_{\pi}$,
\begin{eqnarray}
\chi & \approx & \chi_0 + m_{\pi}^2 \chi_1 + m_{\pi}^4 \chi_2 + m_{\pi}^6 \chi_3,
\nonumber \\
\lambda & \approx  & m_{\pi}^2 \lambda_1 + m_{\pi}^4 \lambda_2 + m_{\pi}^6 \lambda_3.
\label{4.12}
\end{eqnarray}
The Euler-Lagrange equations corresponding to the effective action (\ref{4.3}) can then be solved 
analytically with the NNNLO results ($y=m_{\pi}x$)
\begin{eqnarray}
\chi_0 & = & \arctan {\rm e}^y
\nonumber \\
\lambda_1 & = & - \frac{1}{4} \frac{1}{\cosh^2 y}
\nonumber \\
\chi_1 & = & \frac{1}{16} \frac{\sinh y}{\cosh^2 y} 
\nonumber \\
\lambda_2 & = & - \frac{1}{96} \frac{10 \cosh^2 y - 13}{\cosh^4 y}
\label{4.13} \\
\chi_2 & = & - \frac{1}{2304} \frac{\sinh y (11 \cosh^2 y - 26)}{\cosh^4 y}
\nonumber \\
\lambda_3 & = & - \frac{1}{5760}\frac{562 \cosh^4 y - 3090 \cosh^2 y + 2811}{\cosh^6 y}
\nonumber \\
\chi_3 & = & \frac{\sinh y}{1382400} \frac{( 6271 \cosh^4 y + 29588 \cosh^2 y - 26784)}{\cosh^6 y}
\nonumber
\end{eqnarray}
The baryon mass becomes  
\begin{equation}
\frac{M_{\rm B}}{N} = \frac{m_{\pi}}{2\pi} \left( 1 - \frac{1}{36} m_{\pi}^2 + \frac{13}{3600} m_{\pi}^4 - \frac{1193}{705600}m_{\pi}^6 \right)
\label{4.14}
\end{equation}
As the whole winding number of $\chi$ resides in the LO term $\chi_0$, baryon number is always 1/2. Therefore the complex potential
$S-{\rm i}P$ traces out half a turn around the chiral circle. This is confirmed by plotting $S$ and $P$, showing kink-like behavior of  $S$
like in the massless GN model, see Fig.~2. The presence of a non-vanishing $P$ signals that we are dealing with a new kind
of solitonic baryon here which did not show up yet in any other variant of the GN model family.

\begin{figure}
\begin{center}
\epsfig{file=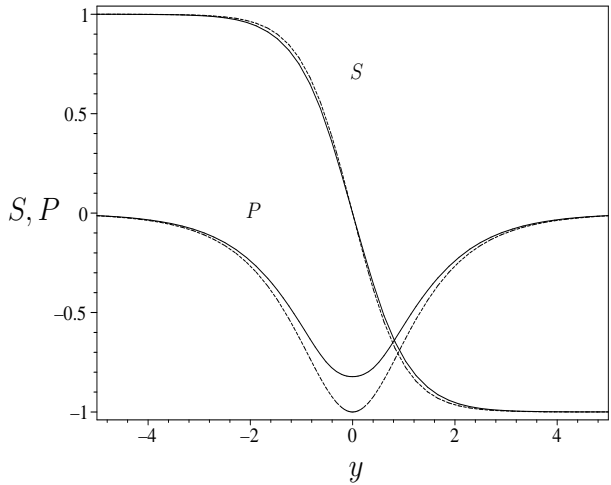,height=8cm,width=6.4cm,angle=270}
\caption{Scalar ($S$) and pseudoscalar ($P$) potentials for baryon in the derivative expansion, $\xi=0.2, m_{\pi}\approx 0.8389$. Dashed curves:
LO (sine-Gordon), solid curves: NNNLO, see Eqs.~(\ref{4.13}).}
\label{fig2}
\end{center}
\end{figure}

\begin{figure}
\begin{center}
\epsfig{file=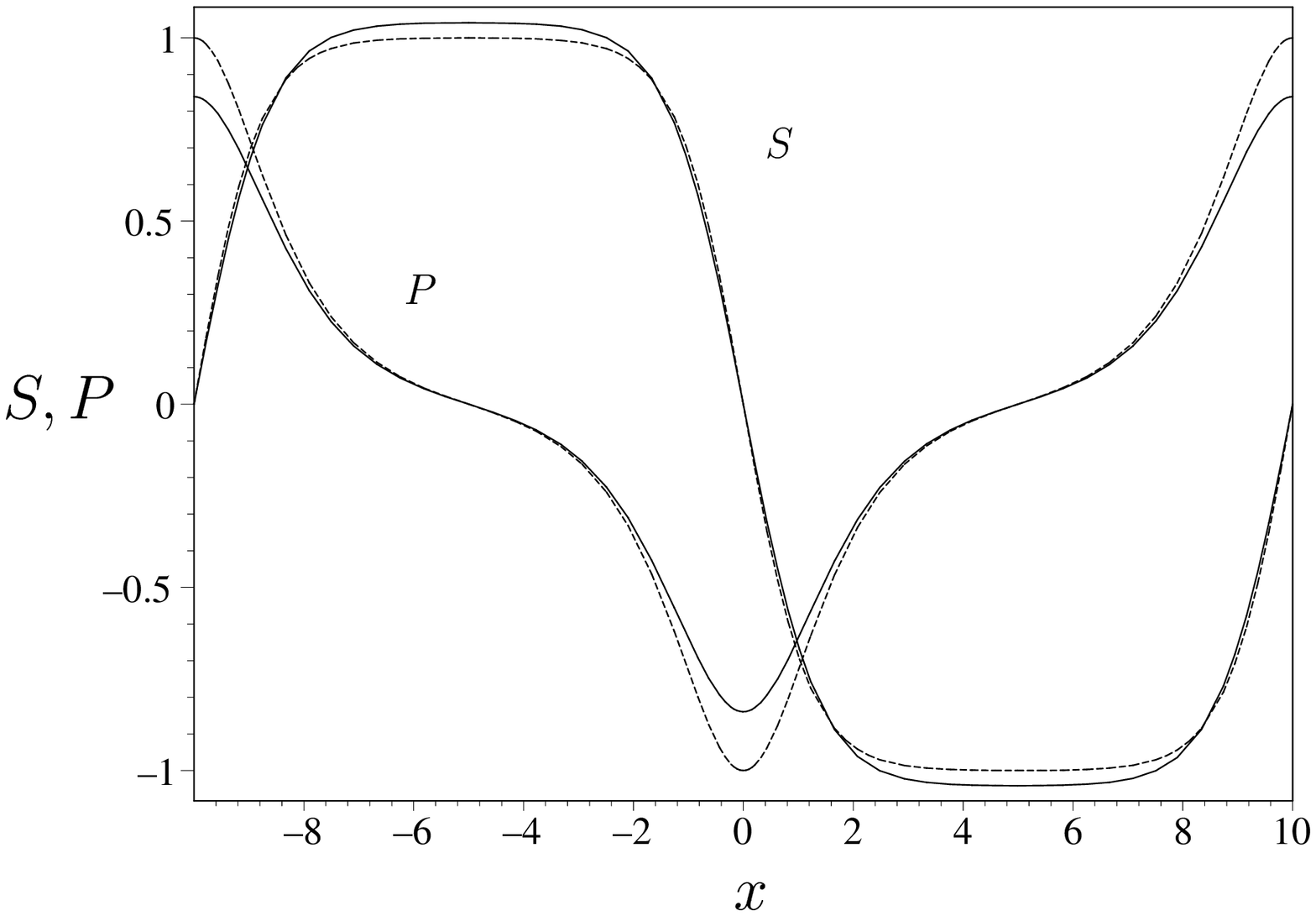,height=8cm,width=6.4cm,angle=270}
\caption{Soliton crystal for generalized GN model, $\xi=0.2, m_{\pi}\approx 0.8389, \rho=0.05$. Dashed curves: LO (sine Gordon),
solid curves: NNLO, see Eqs.~(\ref{4.15}).} 
\label{fig3}
\end{center}
\end{figure}

Let us now turn to periodic solutions of the Euler-Lagrange equations in the derivative expansion. They are expected to 
approximate systematically the ground state of matter at low densities and in the vicinity of the chiral limit $\xi=0$. 
Since the resulting expressions are rather lengthy, we only give them up to NNLO here,
\begin{eqnarray}
\chi_0 & = & \frac{\pi}{4} + \frac{1}{2} {\rm am}
\nonumber \\
\lambda_1 & = & - \frac{1}{4} {\rm cn}^2
\nonumber \\
\chi_1 & = & \left( \frac{\zeta}{24} + \frac{1}{16}\right) {\rm sn} \,{\rm cn} - \frac{\zeta}{24 \kappa^2} {\rm dn}\, {\rm Z}
\nonumber \\
\lambda_2 & = &   \left( \frac{13}{96}- \frac{\zeta}{24} \right) {\rm sn}^4 + \left( \frac{\zeta}{24}- \frac{1+\kappa^2}{24 \kappa^2}
\right) {\rm sn}^2 + \frac{4-\kappa^2}{96 \kappa^2} 
\nonumber \\
& - &  \frac{\zeta}{24 \kappa^2} {\rm sn}\,{\rm cn}\,{\rm dn}\,{\rm Z}
\nonumber \\
\chi_2 & = & \left( \frac{\zeta^3}{576 \kappa^2} + \frac{(\kappa^2-5)\zeta^2}{576 \kappa^2} + \frac{(61+30\kappa^2)\zeta}{2880 \kappa^2} 
\right.
\nonumber \\
& & \left. + \frac{59 \kappa^2-44}{2304 \kappa^2}\right) {\rm sn}\,{\rm cn}
\nonumber \\
& -  &  \left( \frac{\zeta^3}{576 \kappa^4} + \frac{(\kappa^2-3)\zeta^2}{288 \kappa^4} + \frac{(61+30 \kappa^2)\zeta}{2880 \kappa^4} \right){\rm dn}
\,{\rm Z} 
\nonumber \\
& - &  \left( \frac{13}{1152} + \frac{\zeta}{96} + \frac{\zeta^2}{576} \right) {\rm sn}^3{\rm cn}  - \frac{\zeta^2}{576 \kappa^2} {\rm sn}\,{\rm cn} \,{\rm Z}^2     
\nonumber \\
& + & \left( \frac{\zeta^2}{288 \kappa^2} +  \frac{\zeta}{96 \kappa^2} \right) {\rm dn}\, {\rm sn}^2 {\rm Z} 
\label{4.15}
\end{eqnarray}
Here, 
\begin{equation}
\zeta = (1- \kappa^2) \frac{\bf K}{\bf E},
\label{4.16}
\end{equation}
${\bf E}, {\bf K}$ are complete elliptic integrals of $\kappa$ and ${\rm am}, {\rm sn}, {\rm cn}, {\rm dn}$ and ${\rm Z}$ are standard Jacobi elliptic
functions with spatial argument 
\begin{equation}
z= \frac{m_{\pi}}{\kappa}x
\label{4.17}
\end{equation}
and elliptic modulus $\kappa$.
The mean density can be simply inferred from the period of the crystal, 
\begin{equation}
\rho = \frac{m_{\pi}}{4 \kappa {\bf K}}.
\label{4.18}
\end{equation}
By way of example, we show in Fig.~\ref{fig3} the scalar and pseudoscalar potentials corresponding to  
$\xi=0.2$ (as in Fig.~\ref{fig2}) and the density $\rho=0.05$. Again the convergence seems to be very good.

Since the derivative expansion is anyway expected to be most useful at low densities, we note the following simplification in the low density limit:
for $\kappa \to 1$, we can use the approximation $\zeta \approx 0$ and keep $\kappa$ only in the arguments of the Jacobi elliptic functions. 
Expressions (\ref{4.15}) then reduce to periodic extensions of the baryon results obtained by simply replacing 
\begin{equation}
\cosh y \to \frac{1}{{\rm cn}(z, \kappa)}, \qquad \sinh y \to \frac{{\rm sn}(z, \kappa)}{{\rm cn}(z, \kappa)}
\label{4.19}
\end{equation}
in Eqs.~(\ref{4.13}). 

Finally, we derive a sum rule for the baryon number of a single baryon, following Ref.~\cite{17}. This will equip us with a way of testing the results from
the derivative expansion. 
Starting point is the divergence of the axial current in the generalized GN model
\begin{eqnarray}
\partial_{\mu} j_5^{\mu} & = &  -2(g^2-G^2) \bar{\psi}\psi \, \bar{\psi} {\rm i}\gamma_5 \psi
\nonumber \\
& = & 2 \left( S \bar{\psi} {\rm i} \gamma_5 \psi - P \bar{\psi}\psi \right)
\nonumber \\
& = & -2N \left(\frac{1}{NG^2}-\frac{1}{Ng^2}\right) SP
\nonumber \\
& = & - \frac{2N\xi}{\pi} SP,
\label{4.20}
\end{eqnarray}
where we have taken a ground state expectation value and used large $N$ factorization.
Owing to the properties
\begin{equation}
j_5^0 = j^1, \qquad j_5^1 = j^0
\label{4.21}
\end{equation}
specific for 1+1 dimensions, 
we get for stationary states 
\begin{equation}
\partial_1 \rho(x) = - \frac{2 N \xi}{\pi} S(x) P(x).
\label{4.22}
\end{equation}
Twofold integration for the baryon case then leads to a sum rule relating baryon number directly to an integral over the HF potentials $S,P$,
\begin{eqnarray}
\rho(x) & = & - \frac{2N \xi}{\pi} \int_{-\infty}^x {\rm d}x' S(x') P(x')
\label{4.23} \\
\frac{1}{2} & = &  - \frac{2 \xi}{\pi} \int_{-\infty}^{\infty} {\rm d}x \int_{-\infty}^x {\rm d}x' S(x')P(x')
\nonumber \\
& = & \frac{2\xi}{\pi} \int_{-\infty}^{\infty} {\rm d}x x S(x) P(x).
\label{4.24}
\end{eqnarray}
In the last step, partial integration was used. Inserting the results for $S,P$ from the baryon, i.e.,
\begin{eqnarray}
S & = & + (1+\lambda)\cos (2\chi),
\nonumber \\
P & = & -(1+\lambda)\sin (2\chi),
\label{4.25}
\end{eqnarray} 
with $\chi, \lambda$ from Eqs.~(\ref{4.13}),
we find that the sum rule (\ref{4.24}) is only violated at ${\rm O}(m_{\pi}^8)$.
This is a good independent test of a considerable amount of algebra behind the derivative expansion.

\section{Phase diagram near the NJL$_2$ tricritical point ($\xi=0$)}\label{sect5}

\begin{figure}
\begin{center}
\epsfig{file=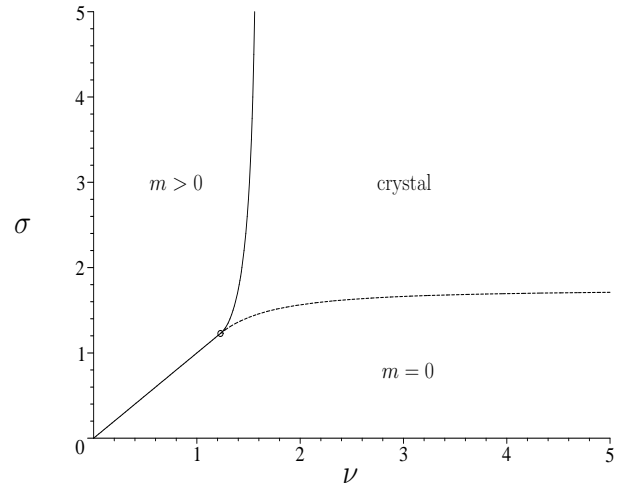,height=8cm,width=6.4cm,angle=270}
\caption{Rescaled phase diagram near the tricritical point of the NJL$_2$ model. 
Straight line: 2nd order phase boundary, Eq.~(\ref{5.12}). 
Dashed curve: 2nd order phase boundary, Eq.~(\ref{5.17}).
Solid curve: 1st order phase boundary, numerical calculation.
The 3 critical curves meet at the tricritical point $\sigma_t=\nu_t=\sqrt{a}/2$.
The parameter $\xi$ has been eliminated by the choice of variables, see Eq.~(\ref{5.7}).}
\label{fig4}
\end{center}
\end{figure}

We start our investigation of the phase diagram of the generalized GN model by zooming in onto the tricritical point at $\xi=0$, i.e., of the 
NJL$_2$ model. In Ref.~\cite{18} it was shown that this region is well suited for the derivative expansion, which here leads to
a (microscopic) Ginzburg-Landau type theory. In that work, chiral symmetry was broken as usual by means of a bare fermion mass term.
Here instead we break it by choosing two slightly different coupling constants in the scalar and pseudoscalar channels. The central quantity of interest
is the grand canonical potential which differs in these two cases only by the double counting correction. Since the latter is independent of temperature and
chemical potential,
the situation is very similar to the one in the preceding section. Once again we can take over the effective action from the literature about the massive
NJL$_2$ model \cite{18}. The only necessary modification is
to replace the double counting correction term coming from the bare mass by the one proportional to $\xi$, cf. Eq.~(\ref{4.1}). 
For the present purpose, it is advantageous to combine the HF potentials $S,P$ into one complex field $\phi=S-{\rm i}P$.
The result for the grand canonical potential density to the order needed here (dropping a field independent part) then becomes
\begin{equation}
\Psi_{\rm eff}= \alpha_2 |\phi|^2 + \alpha_3 \Im (\phi \phi'\,^*) + \alpha_4 \left(|\phi|^4+|\phi'|^2\right)
+ \frac{\xi}{2\pi} (\Im \phi)^2
\label{5.1}
\end{equation}
with 
\begin{eqnarray}
\alpha_2 & = & \frac{1}{2\pi} \left[\ln (4\pi T) + \Re \Psi(z) \right]
\nonumber \\
\alpha_3 & = & - \frac{1}{8\pi^2T}\Im \Psi^{(1)}(z)
\nonumber \\
\alpha_4 & = & - \frac{1}{64 \pi^3 T^2} \Re \Psi^{(2)}(z)
\label{5.2}
\end{eqnarray}
and 
\begin{equation}
z=\frac{1}{2}+ \frac{{\rm i} \mu}{2\pi T}.
\label{5.3}
\end{equation}
We denote the digamma and polygamma functions as
\begin{equation}
\Psi(z) = \frac{\rm d}{{\rm d}z} \ln \Gamma(z), \qquad \Psi^{(n)}(z) = \frac{{\rm d}^n}{{\rm d}z^n}\Psi(z).
\label{5.3a}
\end{equation}
In the chiral limit ($\xi=0$), the tricritical point is located at
\begin{equation}
\mu_t = 0, \quad T_t = T_c = \frac{{\rm e}^{\rm C}}{\pi}
\label{5.4}
\end{equation}
with Euler's constant ${\rm C} \approx 0.577216$. Following Ref.~\cite{18}, we expand the 
coefficients (\ref{5.2}) of the GL effective action around the tricritical point (\ref{5.4}),
\begin{eqnarray}
\alpha_2 & \approx & \frac{7}{8\pi} \zeta(3) {\rm e}^{-2{\rm C}}\mu^2 - \frac{1}{2} {\rm e}^{-{\rm C}}\tau^2
\nonumber \\
\alpha_3 & \approx & \frac{7}{8\pi} \zeta(3) {\rm e}^{-2{\rm C}}\mu
\nonumber \\
\alpha_4 & \approx &  \frac{7}{32 \pi} \zeta(3) {\rm e}^{-2 {\rm C}}
\label{5.5}
\end{eqnarray}
with $\tau=\sqrt{T_c-T}$. The $\xi$-dependence can now be removed as follows. Rescaling
the field and the coordiante according to 
\begin{eqnarray}
\phi(x) & = & \xi^{1/2} \varphi(u), \qquad u=\xi^{1/2}x
\nonumber \\
\phi'(x) & = & \xi \dot{\varphi}(u), \qquad \phi''(x)\ = \ \xi^{3/2} \ddot{\varphi}(u)
\label{5.6}
\end{eqnarray} 
and introducing rescaled thermodynamic variables
\begin{equation}
\nu = \frac{2\mu}{\xi^{1/2}}, \qquad \sigma = \sqrt{\frac{a}{T_c}}\frac{\tau}{\xi^{1/2}}
\label{5.7}
\end{equation}
with the constant
\begin{equation}
a=\frac{16 {\rm e}^{2{\rm C}}}{7\zeta(3)} \approx 6.03198,
\label{5.8}
\end{equation}
the reduced grand canonical potential density
\begin{equation}
\tilde{\Psi}_{\rm eff} = \frac{2\pi a}{\xi^2} \Psi_{\rm eff}
\label{5.9}
\end{equation}
becomes indeed independent of $\xi$,
\begin{eqnarray}
\tilde{\Psi}_{\rm eff} & = & |\dot{\varphi}|^2-{\rm i}\nu (\varphi \dot{\varphi}^*-\dot{\varphi}\varphi^*)+(\nu^2-\sigma^2)|\varphi|^2+|\varphi|^4
\nonumber \\
& &  -\frac{a}{4}(\varphi-\varphi^*)^2.
\label{5.10}
\end{eqnarray}
The Euler-Lagrange equation 
\begin{equation}
\ddot{\varphi}-2{\rm i}\nu \dot{\varphi}+(\sigma^2-\nu^2)\varphi - 2 |\varphi|^2\varphi - \frac{a}{2}(\varphi-\varphi^*) = 0
\label{5.11}
\end{equation}
differs from the complex non-linear Schr\"odinger equation by the term $\sim \varphi^*$. This has prevented us from 
finding the solution in closed analytical form. Let us first determine the expected 2nd order phase boundaries.
The phase boundary between massless and massive homogeneous phases can easily be found by minimizing $\tilde{\Psi}_{\rm eff}$
with the ansatz $\varphi=m$ and setting $m=0$ in the condition for the non-trivial solution. The result in the new coordinates is the straight line
\begin{equation}
\sigma = \nu .
\label{5.12}
\end{equation}
Next consider the phase boundary separating the crystal phase from the chirally restored ($m=0$) homogeneous phase.
Here we use the ansatz (see Sec. IV of Ref.~\cite{18} for the justification) 
\begin{equation}
\varphi = c_0 \cos(qu) + {\rm i}d_0 \sin (qu)
\label{5.13}
\end{equation}
and evaluate the spatial average of $\tilde{\Psi}_{\rm eff}$, keeping only terms up to 2nd order in $c_0,d_0$,
\begin{equation}
\langle \tilde{\Psi}_{\rm eff} \rangle = {\cal M}_{11}c_0^2 +2 {\cal M}_{12}c_0d_0 + {\cal M}_{22} d_0^2,
\label{5.14}
\end{equation}
with
\begin{eqnarray}
{\cal M}_{11} & = &  \frac{1}{2} (q^2+\nu^2-\sigma^2)
\nonumber \\
{\cal M}_{12} & = & -\nu q
\nonumber \\
{\cal M}_{22} & = &  \frac{1}{2} (a+q^2+\nu^2-\sigma^2).
\label{5.15}
\end{eqnarray}
As explained in Ref.~\cite{8}, the phase boundary is now defined by the conditions
\begin{equation}
{\rm det}{\cal M} =0,\qquad \frac{\partial}{\partial q^2} {\rm det}{\cal M} =0 ,
\label{5.16}
\end{equation}
yielding the critical curve
\begin{equation}
\sigma = \frac{\sqrt{a(8\nu^2-a)}}{4\nu} .
\label{5.17}
\end{equation}
The wave number $q$ obeys
\begin{equation}
q= \sqrt{\sigma^2+\nu^2-\frac{a}{2}}.
\label{5.18}
\end{equation}
The tricritical point can be identified with the point of intersection of the two critical
curves (\ref{5.12}) and (\ref{5.17}),
\begin{equation}
\sigma_t = \nu_t =\frac{\sqrt{a}}{2}.
\label{5.19}
\end{equation}
Going back to the original, unscaled variables, this translates into
\begin{eqnarray}
T_t & = & T_c\left(1-\frac{1}{4}\xi \right),
\nonumber \\
\mu_t & = &  \frac{\sqrt{a}}{4} \xi^{1/2}.
\label{5.19a}
\end{eqnarray}
Notice that $q$ vanishes at the tricritical point.
We expect that a third critical line ends at the tricritical point, namely the 1st order phase boundary separating 
the crystal from the massive Fermi gas phase. It has to be determined numerically. To this end, we insert the Fourier 
series ansatz
\begin{equation}
\varphi = \sum_{n} c_n \cos [ (2n+1)qu] + {\rm i} \sum_n d_n \sin [(2n+1)qu]
\label{5.20}
\end{equation}
into Eq.~(\ref{5.10}) and minimize the effective action with respect to the parameters $c_n, d_n$ and $q$. By keeping only 
wave numbers which are odd multiples of $q$, we restrict ourselves to potentials which are antiperiodic over half a period,
\begin{equation}
\varphi(u+\pi/q) = - \varphi(u).
\label{5.21}
\end{equation}    
This kind of shape is indeed favored by the minimization, as was the case for the massless GN model.
It shows that discrete chiral symmetry and translational symmetry are broken down to a discrete combination of the 2 
transformations, namely
\begin{equation}
\psi(x) \to \gamma_5 \psi(x+\pi/q)
\label{5.22}
\end{equation}
from which Eq.~(\ref{5.21}) for bilinears follows.
In practice, we found that it is sufficient to keep $c_0,c_1,d_0,d_1$ in the expansion (\ref{5.20}).
Comparing the reduced grand potential with the one from the homogeneous massive solution, we can locate the
phase boundary. The result of the calculation is shown in Fig.~\ref{fig4} together with the two 2nd order phase
boundaries discussed above. Due to the rescalings, this is a kind of universal phase diagram which contains all information about
the actual phase diagram in the vicinity of the tricritical point at $\xi=0$. By undoing the rescaling we can reconstruct the
phase diagrams for small $\xi$ values in a limited region of the ($\mu,T$) plane. This is shown in Fig.~5.
Here one sees nicely the transition from the behavior
qualitatively familiar from the GN model to the one from the massless NJL$_2$ model. The angle between the two
phase boundaries delimiting the crystal at the tricritical point is consistent with zero, just like in the standard GN model. 

\begin{figure}
\begin{center}
\epsfig{file=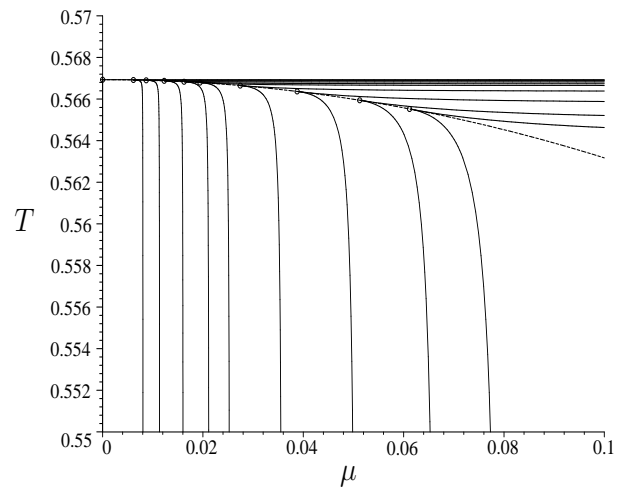,height=8cm,width=6.4cm,angle=270}
\caption{Reconstructed phase diagram of generalized GN model near the tricritical point of NJL$_2$ model for  $\xi=$ 0.0001, 0.0002, 0.0004, 0.0007, 0.001, 0.002,
0.004, 0.007, 0.01, from left to right. All curves are obtained from the ones shown in Fig.~\ref{fig4}, but $\nu,\tau$ values up to $\approx$ 50 are needed for the 
smallest $\xi$ value.}
\label{fig5}
\end{center}
\end{figure}

\section{Exact tricritical behavior from Ginzburg-Landau theory}\label{sect6}

As $\xi$ varies from 0 to $\infty$, the tricritical point of the generalized GN model moves from the NJL$_2$ to the GN tricritical 
point, i.e. from $\mu=0,T=0.5669$ to $\mu=0.6082,T=0.3183$. Since the HF potential $\phi=S-{\rm i}P$ vanishes at the tricritical point and its period
is expected to diverge, the derivative expansion should be sufficient to determine the exact tricritical behavior for all $\xi$.
As a matter of fact, this will enable us to determine analytically the location of the tricritical point as a function of $\xi$. We will also
be interested in the behavior of the phase boundaries in the vicinity of the tricritical point. It turns out that the region of validity
of the GL theory as defined in Eq.~(\ref{5.1}) shrinks rapidly with increasing $\xi$. One of the reasons is the fact that both $\alpha_2$
and $\alpha_4$ vanish at the GN tricritical point, so that it would be necessary to go to higher orders in the derivative expansion for large $\xi$.
To keep the analytical work reasonably simple, we therefore analyze the phase boundaries only for moderate $\xi$ values. 

We start once again from the GL effective action (\ref{5.1}). Consider first the homogeneous phases. The constant ansatz
$\phi  =  m$
yields
\begin{equation}
\Psi_{\rm eff}  =  \alpha_2 m^2 + \alpha_4 m^4.
\label{6.1}
\end{equation}
Minimizing with respect to $m$, we find either $m=0$ or 
\begin{equation}
m = \sqrt{-\frac{\alpha_2}{2\alpha_4}} \qquad (\alpha_2<0).
\label{6.2}
\end{equation}
We thus recover the well known result for the phase boundary between massless and massive Fermi gas phases, namely
\begin{equation}
\alpha_2=0
\label{6.3}
\end{equation}
or, parametrically (parameter $\tilde{\nu}$),
\begin{eqnarray}
T & = & \frac{1}{4\pi} {\rm e}^{- \Re \Psi(z)} \qquad \left(z=\frac{1}{2}+ {\rm i} \frac{\tilde{\nu}}{2\pi}\right),
\nonumber \\
\mu & = & \tilde{\nu} T.
\label{6.4}
\end{eqnarray}
Next consider the 2nd order phase boundary between crystal and massless homogeneous phase. As in Sec.~\ref{sect5}, 
the ansatz
\begin{equation}
\phi = c_0 \cos (Qx) + {\rm i} d_0 \sin (Qx)
\label{6.5}
\end{equation}
is adequate for a continuous phase transition which can be treated in perturbation theory.
The spatial average of the effective action, keeping only quadratic terms 
in ($c_0, d_0$), then becomes 
\begin{equation}
\langle \Psi_{\rm eff} \rangle = {\cal M}_{11} c_0^2+ 2 {\cal M}_{12} c_0 d_0 + {\cal M}_{22} d_0^2
\label{6.6}
\end{equation}
where
\begin{eqnarray}
{\cal M}_{11} & = & \frac{1}{2} \left( \alpha_2 + \alpha_4 Q^2 \right)
\nonumber \\
{\cal M}_{12} & = & - \frac{1}{2} \alpha_3 Q
\nonumber \\
{\cal M}_{22} & = & \frac{1}{2} \left( \alpha_2 + \alpha_4 Q^2 + \frac{\xi}{2\pi} \right).
\label{6.7}
\end{eqnarray}
The 2nd order phase boundary is again defined by
\begin{equation}
{\rm det}{\cal M} = 0, \qquad \frac{\partial}{\partial Q^2}\, {\rm det}{\cal M} = 0
\label{6.8}
\end{equation}
or, equivalently,
\begin{eqnarray}
0 & = & Q^4   + \left(\frac{\xi}{2\pi \alpha_4}  - \left( \frac{\alpha_3}{\alpha_4}\right)^2 + \frac{2 \alpha_2}{\alpha_4} \right)Q^2
+ \frac{\alpha_2 \xi}{2\pi \alpha_4^2} + \frac{\alpha_2^2}{\alpha_4^2} 
\nonumber \\
0 & = & Q^2 + \frac{\xi}{4\pi \alpha_4} + \frac{\alpha_2}{\alpha_4} - \frac{\alpha_3^2}{2 \alpha_4^2} .
\label{6.9}
\end{eqnarray}
These two equations determine $Q$ and the critical curve in the ($\mu,T$) plane.
The tricritical point must lie on this curve and on the curve $\alpha_2=0$. This gives the conditions $Q=0$ and 
\begin{equation}
\xi = \left. \frac{2\pi \alpha_3^2}{\alpha_4} \right|_t,
\label{6.10}
\end{equation}
where the right hand side is to be evaluated at the tricritical point.
Using Eqs.~(\ref{5.2}), we finally arrive at the following parametric representation of the dependence of the tricritical point
($\mu_t,T_t$) on $\xi$ (parameter $\tilde{\nu}_t$), 
\begin{eqnarray}
\xi & = & -  \frac{2 \left[\Im \Psi^{(1)}(z_t)\right]^2}{\Re \Psi^{(2)}(z_t)} \qquad \left( z_t=\frac{1}{2}+ \frac{{\rm i} \tilde{\nu}_t}{2\pi}
\right)
\nonumber \\
T_t & = & \frac{1}{4\pi} {\rm e}^{- \Re \Psi(z_t)}
\nonumber \\
\mu_t & =  & \tilde{\nu}_t T_t.
\label{6.11}
\end{eqnarray}
This result should hold exactly in the generalized GN model, since GL theory becomes rigorous at the tricritical point.
It has the correct limits for $\xi \to 0$ (NJL$_2$) and $\xi \to \infty$ (GN), as follows immediately from the vanishing of $\alpha_3$
and $\alpha_4$, respectively. Moreover, by expanding in $\nu_t$ we recover the asymptotic behavior of ($\mu_t,T_t$) for $\xi \to 0$
found in Sec.~\ref{sect5}, cf. Eq.~(\ref{5.19a}). 

\begin{figure}
\begin{center}
\epsfig{file=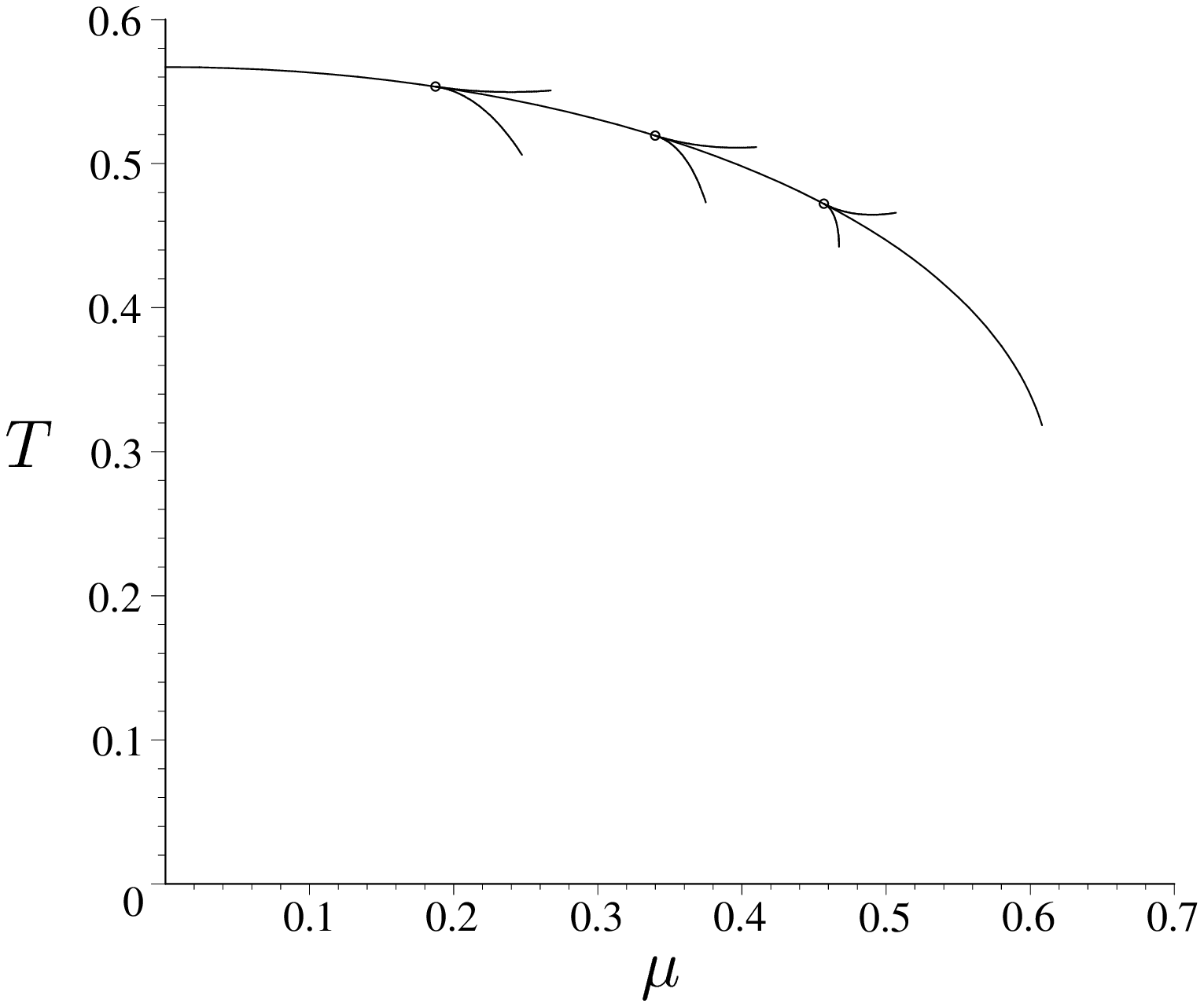,height=8cm,width=6.4cm,angle=270}
\caption{Tricritical behavior from GL theory for $\xi=0.1,0.4,1.0$, from left to right. 
This figure shows how the 1st order and 2nd order phase boundaries merge in a cusp at the tricritical point.}
\label{fig6}
\end{center}
\end{figure}

We now determine the shape of the phase boundaries near the tricritical point for finite $\xi$ values. To this end, 
we measure chemical potential and temperature from the tricritical
point (at fixed $\xi$),
\begin{eqnarray}
\mu & = & \mu_{t} + \delta,
\nonumber \\
T & = & T_t + \tau.
\label{6.12}
\end{eqnarray}
We then rotate the coordinate frame in the ($\delta, \tau$) plane such that the new axes are tangential and normal to the 
homogeneous phase boundary $\alpha_2=0$,
\begin{equation}
\left( \begin{array}{c} \delta \\ \tau \end{array} \right) = \left( \begin{array}{rr} \cos \theta & - \sin \theta \\ \sin \theta &
\cos \theta \end{array} \right)\left( \begin{array}{c} \sigma \\ \eta \end{array} \right)
\label{6.13}
\end{equation}
with
\begin{equation}
\sin \theta = \frac{I_1}{\Omega}, \qquad \cos \theta  =  \frac{\zeta}{\Omega}
\label{6.14}
\end{equation}
We have defined
\begin{equation}
\zeta = 2\pi+ \tilde{\nu}_t I_1, \quad \Omega = \sqrt{ I_1^2+\zeta^2}, \quad I_1=\Im \Psi^{(1)}(z_t).
\label{6.15}
\end{equation}
Due to the cusp, the phase boundaries lie in the region around the tricritical point where 
\begin{equation}
\sigma \sim \varepsilon, \qquad \eta \sim \varepsilon^2.
\label{6.16}
\end{equation}
In this region, the Taylor expansion
\begin{eqnarray}
\alpha_2 & = & a_{22} \varepsilon^2 + ...
\nonumber \\
\alpha_3  & = & a_{30}+ a_{31} \varepsilon + a_{32} \varepsilon^2 + ...
\nonumber \\
\alpha_4  & = &  a_{40} + a_{41}\varepsilon + a_{42} \varepsilon^2 + ...
\label{6.17}
\end{eqnarray}
holds with calculable coefficients given in the appendix.
We first determine the shape of the 2nd order phase boundary from (\ref{6.9},\ref{6.10}) and (\ref{6.17}). 
To leading order in $\varepsilon$, we find
\begin{equation}
Q^2 = \frac{a_{30}(2 a_{31}a_{40}-a_{30}a_{41})}{2 a_{40}^3} \varepsilon
\label{6.18}
\end{equation}
and the following condition for the phase boundary,
\begin{equation}
0 = 4 a_{30}a_{31}a_{40}a_{41} + 4 a_{22}a_{40}^3-4 a_{31}^2 a_{40}^2 -a_{30}^2 a_{41}^2.
\label{6.19}
\end{equation}
We can also determine the ratio $d_0/c_0$ of imaginary to real amplitudes, 
\begin{equation}
\frac{d_0}{c_0} = \sqrt{\frac{2 a_{31}a_{40}-a_{30}a_{41}}{2 a_{30}a_{40}}}\sqrt{\varepsilon}.
\label{6.20}
\end{equation}
Computing the 1st order phase boundary is the most complicated task.
Let us decompose $\phi$ into real and imaginary parts and assume the following LO behavior in $\varepsilon$,
\begin{eqnarray}
\phi & = & F+{\rm i} G
\nonumber \\
y & = & \varepsilon^{1/2} x
\nonumber \\
F(x) & = & \varepsilon F_0(y), \qquad F'(x) = \varepsilon^{3/2} \dot{F}_0(y)
\nonumber \\ 
G(x) & = & \varepsilon^{3/2} G_0(y), \ \   G'(x) = \varepsilon^2 \dot{G}_0(y)
\label{6.21}
\end{eqnarray}
These assumptions will be justified a posteriori once we have constructed a consistent solution.
We then get
\begin{eqnarray}
\Psi_{\rm eff} & = & \left(-a_{30} F_0 \dot{G}_0+ a_{30} G_0 \dot{F}_0+ \frac{a_{30}^2}{a_{40}} G_0^2 + a_{40} \dot{F}_0^2\right) \varepsilon^3
\nonumber \\
& & + \left( a_{22} F_0^2- a_{31} F_0 \dot{G}_0 + a_{31} G_0 \dot{F}_0 + a_{40} F_0^4 \right.
\nonumber \\
& & \left. +\, a_{40} \dot{G}_0^2 + a_{41}\dot{F}_0^2\right) \varepsilon^4
\label{6.22}
\end{eqnarray}
$G_0$ can be eliminated as follows: Vary the O($\varepsilon^3$) term with respect to $G_0$, find the condition
\begin{equation}
G_0 = - \frac{a_{40}}{a_{30}} \dot{F}_0.
\label{6.23}
\end{equation}
If we insert this relation into Eq.~(\ref{6.22}), the $\varepsilon^3$ term disappears after a partial integration and we are left with
\begin{equation}
\Psi_{\rm eff}  =  \frac{a_{40}^3}{a_{30}^2} \ddot{F}_0^2- \frac{2 a_{31}a_{40}-a_{30}a_{41}}{a_{30}} \dot{F}_0^2 + a_{40} F_0^4 
 + a_{22} F_0^2 .
\label{6.24}
\end{equation}
Here we have set the formal expansion parameter $\varepsilon=1$ since it is not needed anymore.
The coefficients may be simplified by rescaling,
\begin{equation}
F_0(y) = \lambda f(\chi y).
\label{6.25}
\end{equation}
The choice 
\begin{eqnarray}
\lambda & = &  \frac{2 a_{31}a_{40}-a_{30}a_{41}}{a_{40}^2}
\nonumber \\
\chi & = & \sqrt{\frac{a_{30}\lambda}{a_{40}}}
\label{6.26}
\end{eqnarray}
then yields the simpler expression
\begin{equation}
\Psi_{\rm eff}  =  {\cal N} \left[ (f'')^2 - (f')^2 +f^4 + \kappa f^2 \right]
\label{6.27}
\end{equation}
with only two residual parameters
\begin{eqnarray}
{\cal N} & = & a_{40} \lambda^4 ,
\nonumber \\
\kappa &=& \frac{a_{22}}{a_{40}} \frac{1}{\lambda^2}.
\label{6.28}
\end{eqnarray}
Now we focus on the reduced effective action
\begin{equation}
\frac{\Psi_{\rm eff}}{\cal N} = \psi_{\rm eff} = (f'')^2 - (f')^2 +f^4 + \kappa f^2.
\label{6.29}
\end{equation}
As we have not been able to solve the Euler-Lagrange equation
\begin{equation}
f^{IV} + f'' + 2 f^3 + \kappa f =0
\label{6.30}
\end{equation}
analytically, we minimize the reduced effective action with the Fourier series ansatz
\begin{equation}
f(z) = \sum_{n=0}^{n_{\rm max}} c_n \cos \left[ (2n+1)qz \right].
\label{6.31}
\end{equation}
Provided we keep only one term in the sum ($n_{\rm max} = 0$), everything can be worked out analytically with the result
\begin{eqnarray}
c_0 & = & \sqrt{\frac{1-4\kappa}{6}},
\nonumber \\
q & = & \frac{1}{\sqrt{2}}.
\label{6.32}
\end{eqnarray}
The (spatially averaged) reduced effective action in this approximation is given by
\begin{equation}
\langle \psi_{\rm eff} \rangle = -\frac{1}{96} (1-4 \kappa)^2.
\label{6.33}
\end{equation}
The 2nd order phase boundary is obtained from $ \langle \psi_{\rm eff} \rangle = 0 $ and assumes the simple form
\begin{equation}
\kappa = \frac{1}{4}.
\label{6.34}
\end{equation}
The homogeneous, massive solution in the rescaled model is characterized by
\begin{equation}
q=0, \qquad c_0 =  \sqrt{- \frac{\kappa}{2}}
\label{6.35} 
\end{equation}
and has the reduced action
\begin{equation}
\psi_{\rm hom} = - \frac{1}{4} \kappa^2.
\label{6.36}
\end{equation}
The 1st order phase boundary then follows from the condition $\langle \psi_{\rm eff}\rangle = \psi_{\rm hom}$ or
\begin{equation}
\kappa =  - \frac{1}{2} - \frac{\sqrt{6}}{4} . 
\label{6.37}
\end{equation}
Eq.~(\ref{6.37}) defines the 1st order phase boundary in the tricritical region. Since the final formulae for all coefficients and phase boundaries
are quite complicated, we have collected them in the appendix. These results have been used to draw the tricritical behavior
for 3 values of $\xi$ as shown in Fig.~\ref{fig6}.

Truncating the Fourier series (\ref{6.31}) after a single term may seem too crude an approximation. Actually, if we keep more terms and 
minimize the effective action numerically, we get results which are almost indistinguishable on our plot.
To illustrate this point, we take the simpler case where we move along the
homogeneous phase boundary.
According to Eq.~(\ref{6.28}), $\kappa=0$ and the reduced effective action (\ref{6.29})
becomes 
\begin{equation}
\psi_{\rm eff} = (f'')^2 - (f')^2 +f^4 .
\label{6.38}
\end{equation}
Let us minimize this action using the Fourier ansatz (\ref{6.31}).
For $n_{\rm max}=0$, we find the analytical result from above,
\begin{eqnarray}
c_0 & = & \frac{1}{\sqrt{6}} \ = \ 0.4082482904,
\nonumber \\
q & = & \frac{1}{\sqrt{2}} \ = \ 0.7071067812.
\label{6.39}
\end{eqnarray}
For larger values of $n_{\rm max}$, the minimization has to be done numerically. The following result for $n_{\rm max}=3$ is 
sufficient for all practical purposes,
\begin{eqnarray}
c_0 & = & +0.4092971855
\nonumber \\
c_1 & = & -0.0021185033
\nonumber \\
c_2 & = & +0.0000036861
\nonumber \\
c_3 & = & -0.0000000064
\nonumber \\
q & = & + 0.7064259383
\label{6.40}
\end{eqnarray}
Due to the rapid convergence of the Fourier series, the lowest order approximation ($n_{\rm max}=0$) to $f(z)$ is 
already very close to the full result. Likewise, a calculation of the 
spatially averaged effective action, 
\begin{eqnarray}
\langle \psi_{\rm eff} \rangle & = & - 0.0104166667 \qquad (n_{\rm max}=0)
\nonumber \\
\langle \psi_{\rm eff} \rangle & = & - 0.0104526283 \qquad (n_{\rm max}=3)
\label{6.41}
\end{eqnarray}
confirms the excellent convergence. 

\section{Full phase diagram}\label{sect7}

So far, we have discussed only those results about the generalized GN model that could be obtained analytically, or at least with a minimal numerical effort.
For the sake of completeness we have also determined the full phase diagram with the help of the HF approach for a number
of values of the parameter $\xi$, interpolating between the well-known GN and NJL$_2$ phase diagrams.
As is clear from the previous sections, for each $\xi$ one needs to determine three phase boundaries meeting at the tricritical point:
\begin{itemize}
\item  The 2nd order critical line separating massless and massive homogeneous phases, identical to the corresponding phase boundary in the
original phase diagram of the GN model \cite{19}. This phase boundary has already been discussed in Sec.~\ref{sect6} and is given analytically
by Eqs.~(\ref{6.3},\ref{6.4}). In our case, it connects the NJL$_2$ critical point to the critical point for a given value of $\xi$, Eq.~(\ref{6.11}).
\item  The 2nd order phase boundary separating the soliton crystal from the massless homogeneous phase which can be determined 
perturbatively (i.e., treating the potentials $S,P$ in the Dirac-HF equation in 2nd order perturbation theory).
The numerical work here amounts to one-dimensional numerical integrations and solution of transcendental equations and can be done easily to
any desired accuracy. Moreover, an asymptotic expression for large chemical potential will be given in closed analytical form. 
\item A 1st order phase boundary between crystal phase and massive Fermi gas which requires a full numerical HF calculation.
Since the technique has been set up previously in a study of the massive NJL$_2$ model and is described in detail in Ref.~\cite{8}, we shall be
very brief here and merely show the final results.
\end{itemize}
Consider the perturbative phase boundary between crystal and massless Fermi gas first. The calculation is similar to the corresponding one for 
the massive NJL$_2$ model \cite{8}, except that we may set $m=0$ right away. Introducing the Fourier
components $S_1,P_1$ of the HF potentials via
\begin{equation}
S(x) = 2 S_1 \cos\left(2 p_f x\right), \quad P(x)=2 P_1 \sin\left(2 p_f x\right)
\label{7.1}
\end{equation}
where the Fermi momentum $p_f$ is related to the mean fermion density as
\begin{equation}
\rho = \frac{1}{a} = \frac{p_f}{\pi},
\label{7.2}
\end{equation} 
the single particle energies in 2nd order perturbation theory read
\begin{equation}
E_{\eta,p}  =   p + \frac{(S_1-P_1)^2}{2(p+p_f)} + \frac{(S_1+P_1)^2}{2(p-p_f)} \quad (\eta p>0)
\nonumber 
\end{equation}
\begin{equation}
E_{\eta,p}  =  - p - \frac{(S_1+P_1)^2}{2(p+p_f)} - \frac{(S_1-P_1)^2}{2(p-p_f)} \quad (\eta p<0)
\label{7.3}
\end{equation}
The correction to the single particle contribution of the grand canonical potential density 
is then given by
\begin{equation}
\delta \Psi_{\rm s.p.} = {\rm P.V.}\int_0^{\Lambda/2}{\rm d}p \left( f_1+f_2+f_3+f_4 \right)
\label{7.4}
\end{equation}
with
\begin{eqnarray}
f_1 & = & - \frac{p(S_1^2+P_1^2)}{\pi (p^2-p_f^2)}, \quad
f_2 \ = \ \frac{2 p_f P_1 S_1}{\pi (p^2-p_f^2)}
\label{7.5} \\
f_3 & = & \frac{p(S_1^2+P_1^2)}{\pi (p^2-p_f^2)} \left( \frac{1}{1+{\rm e}^{\beta(p-\mu)}} +  \frac{1}{1+{\rm e}^{\beta(p+\mu)}} \right)
\nonumber \\
f_4 & = &  \frac{2p_f P_1 S_1}{\pi (p^2-p_f^2)} \left( \frac{1}{1+{\rm e}^{\beta(p-\mu)}} -  \frac{1}{1+{\rm e}^{\beta(p+\mu)}} \right)
\nonumber
\end{eqnarray}
As in any HF calculation it has to be supplemented by the 
double counting correction,
\begin{equation}
\delta \Psi_{\rm d.c.} = \frac{1}{\pi} (S_1^2+P_1^2)\ln \Lambda + \frac{\xi}{\pi} P_1^2.
\label{7.6}
\end{equation}
Carrying out the principal value integrals involving $f_1,f_2$ analytically, we arrive at the 
finite expression for the sum of (\ref{7.5}) and (\ref{7.6})
\begin{equation}
\delta \Psi  =  \frac{1}{\pi} (S_1^2+P_1^2) \ln (2p_f) + \frac{\xi}{\pi} P_1^2 \, {\rm P.V.}
\int_0^{\infty} {\rm d}p (f_3+f_4)
\label{7.7}
\end{equation}
From here on, we can proceed in the same manner as in the previous sections, i.e., we set
\begin{equation}
\delta \Psi = {\cal M}_{11} S_1^2+ 2 {\cal M}_{12} S_1 P_1 + {\cal M}_{22} P_1^2
\label{7.8}
\end{equation}
and solve the equations
\begin{equation}
{\rm det}{\cal M} = 0, \qquad \frac{\partial}{\partial p_f} {\rm det}{\cal M} = 0
\label{7.9}
\end{equation}
numerically. Further simplifications occur at large $\mu$ where the asymptotic behavior of the phase boundary
can be determined analytically.
Once again we take over the corresponding formula from the massive NJL$_2$ model \cite{8}, merely 
modifying the double counting correction and dropping the $S_0(=m)$ piece. Setting $S_1=X+y/2, P_1=X-y/2$,
we then get
\begin{eqnarray}
\Psi_{\rm eff} &=& \frac{2 X^2}{\pi} \ln (4 p_f) + \frac{y^2}{4\pi} \left( \ln(y^2)-1 \right)
 + \frac{\xi}{\pi} \left(X-\frac{y}{2}\right)^2
\nonumber \\
& & - \frac{2}{\beta \pi} \int_0^{\infty} {\rm d}p \ln \left( 1+ {\rm e}^{-\beta \sqrt{p^2+y^2}}\right).
\label{7.10}
\end{eqnarray}
Minimization with respect to $X$ yields
\begin{equation}
X= \frac{\xi y}{4 \ln (4 p_f)+2 \xi}.
\label{7.11}
\end{equation}
Minimization with respect to $y$ gives the condition
\begin{eqnarray}
0 & = &  2 \int_0^{\infty} {\rm d}p \frac{1}{\sqrt{p^2+y^2}\left(1+{\rm e}^{\beta \sqrt{p^2+y^2}}\right)}
\nonumber \\
& & + \ln y + \frac{\xi \ln (4 p_f)}{\xi+ 2 \ln (4 p_f)} .
\label{7.12}
\end{eqnarray}
Expanding the integral in (\ref{7.12}) for small $y$ \cite{20},
\begin{equation}
0  =    \ln y + \frac{\xi \ln (4 p_f)}{\xi+ 2 \ln (4 p_f)}  - \ln \frac{\beta y}{\pi} - {\rm C} + {\rm O}(y^2),
\label{7.13}
\end{equation}
the asymptotic form of the phase boundary is finally given by the expression
($\mu \approx p_f$),
\begin{equation}
T_{\rm crit} = \frac{{\rm e}^{\rm C}}{\pi} {\rm e}^{-K}, \qquad K=\frac{\xi \ln (4\mu)}{\xi + 2 \ln (4\mu)}.
\label{7.14}
\end{equation}
$X$ in Eq.~(\ref{7.11}) interpolates between 0 (NJL$_2$) and $y/2$ (GN) for $\xi=0...\infty$. Likewise, $T_{\rm crit}$ smoothly
interpolates between the known results for the NJL$_2$ and GN model, respectively.

\begin{figure}
\begin{center}
\epsfig{file=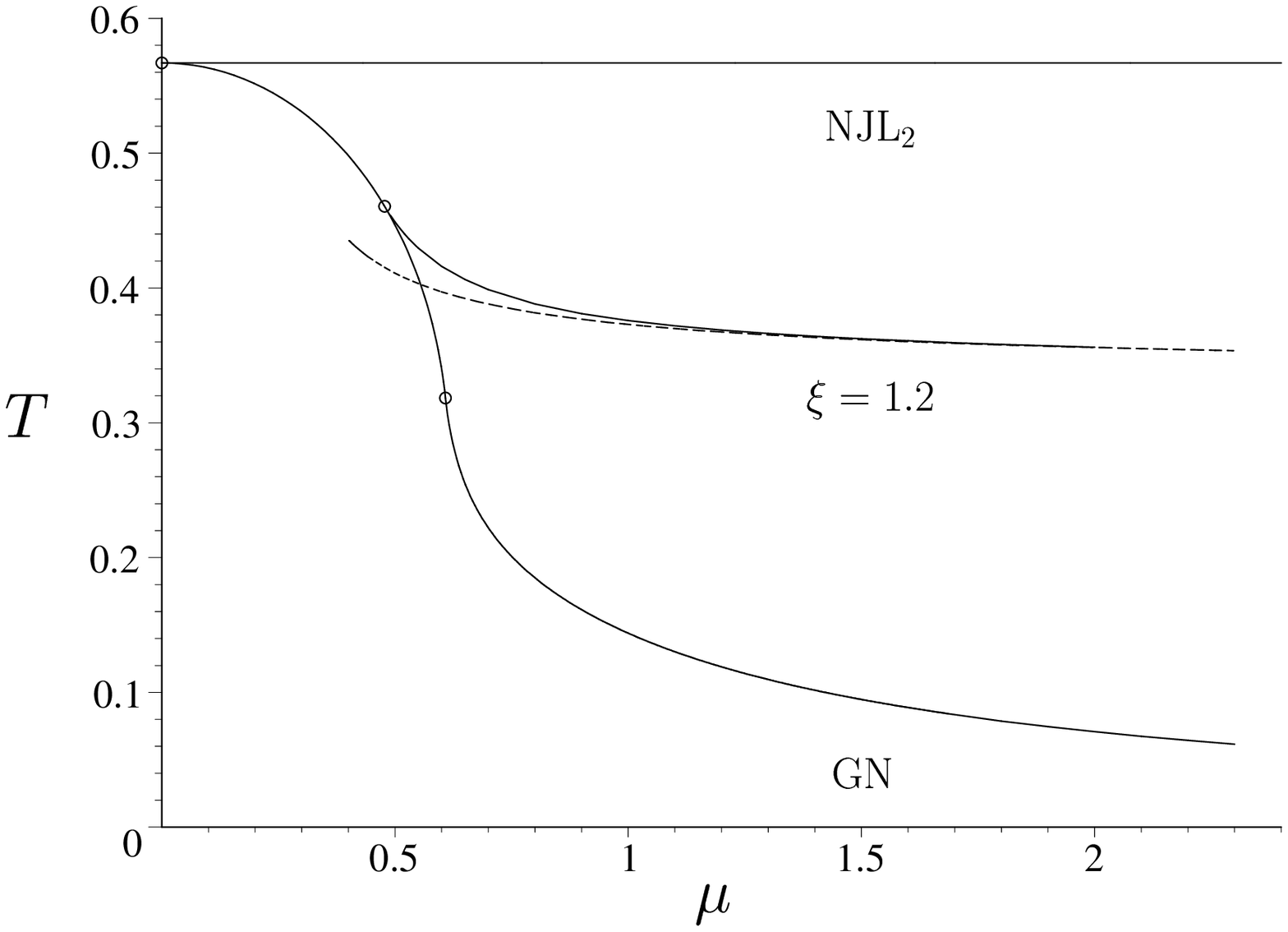,height=8cm,width=6.4cm,angle=270}
\caption{Perturbative 2nd order phase boundary separating the crystal from the chirally restored homogeneous phase at $\xi=1.2$.
Also shown are the corresponding critical lines for the NJL$_2$ model ($\xi=0$) and the GN model ($\xi=\infty$).
Dashed curve: asymptotic expression, Eq.~(\ref{7.14}). The open circles are the tricritical points for all 3 cases.}
\label{fig7}
\end{center}
\end{figure}

\begin{figure}
\begin{center}
\epsfig{file=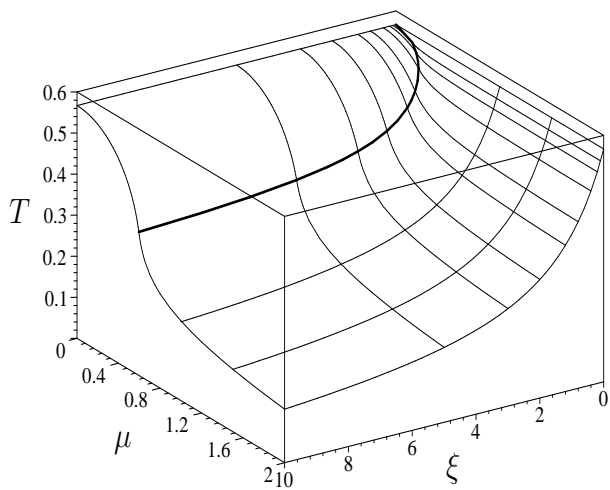,height=8cm,width=6.4cm,angle=270}
\caption{Like Fig.~\ref{fig7}, 3d plot  for several values of $\xi$ ($\xi$ = 0, 0.1, 0.2, 0.4, 0.8, 1.2, 2.0, 3.0, 5.0, 10.0). Fat curve: Tricritical line.
Three curves at constant $\mu$ ($\mu$ = 1.0, 1.5, 2.0): asymptotic expression,  Eq.~(\ref{7.14}).}
\label{fig8}
\end{center}
\end{figure}

In Fig.~\ref{fig7} we show by way of example the perturbative phase boundary at $\xi=1.2$, together with the NJL$_2$ ($\xi=0$) 
and GN ($\xi \to \infty$) model phase boundaries. The asymptotic expression (\ref{7.14}) is shown as the dashed curve and 
only deviates from the full result below $\mu \approx 1$. 
Fig.~\ref{fig8} represents a 3d plot of the perturbative phase
boundary for 10 values of $\xi$ ranging from 0 to 10. The thick line is the tricritcial curve. We have also drawn asymptotic
behavior according to Eq.~(\ref{7.14}) for 3 moderate values of $\mu$ to demonstrate how well this simple formula catches the perturbative 
critical sheet for all values of $\xi$, starting from $\mu \approx 1$.

Still missing in Fig.~\ref{fig8} is the critical sheet separating the crystal from the massive Fermi gas. We recall that this phase transition
is of 2nd order in the GN model, non-existing in the massless NJL$_2$ model and of 1st order in the massive NJL$_2$ model. We find that it
is of 1st order in the generalized GN model for all values of $\xi$, so that apparently the phase transition becomes continuous only in the
GN limit $\xi \to \infty$. Hence there is
no way of determining the critical sheet perturbatively and we need a full thermal HF calculation. Fortunately, this
can be done using the techniques which have recently been developed for the massive NJL$_2$ model \cite{8}. As a matter of fact, 
all what is needed is a trivial modification of the double counting correction.  
We therefore refer to Ref.~\cite{8} for more technical details and immediately pass on to the results.

\begin{figure}
\begin{center}
\epsfig{file=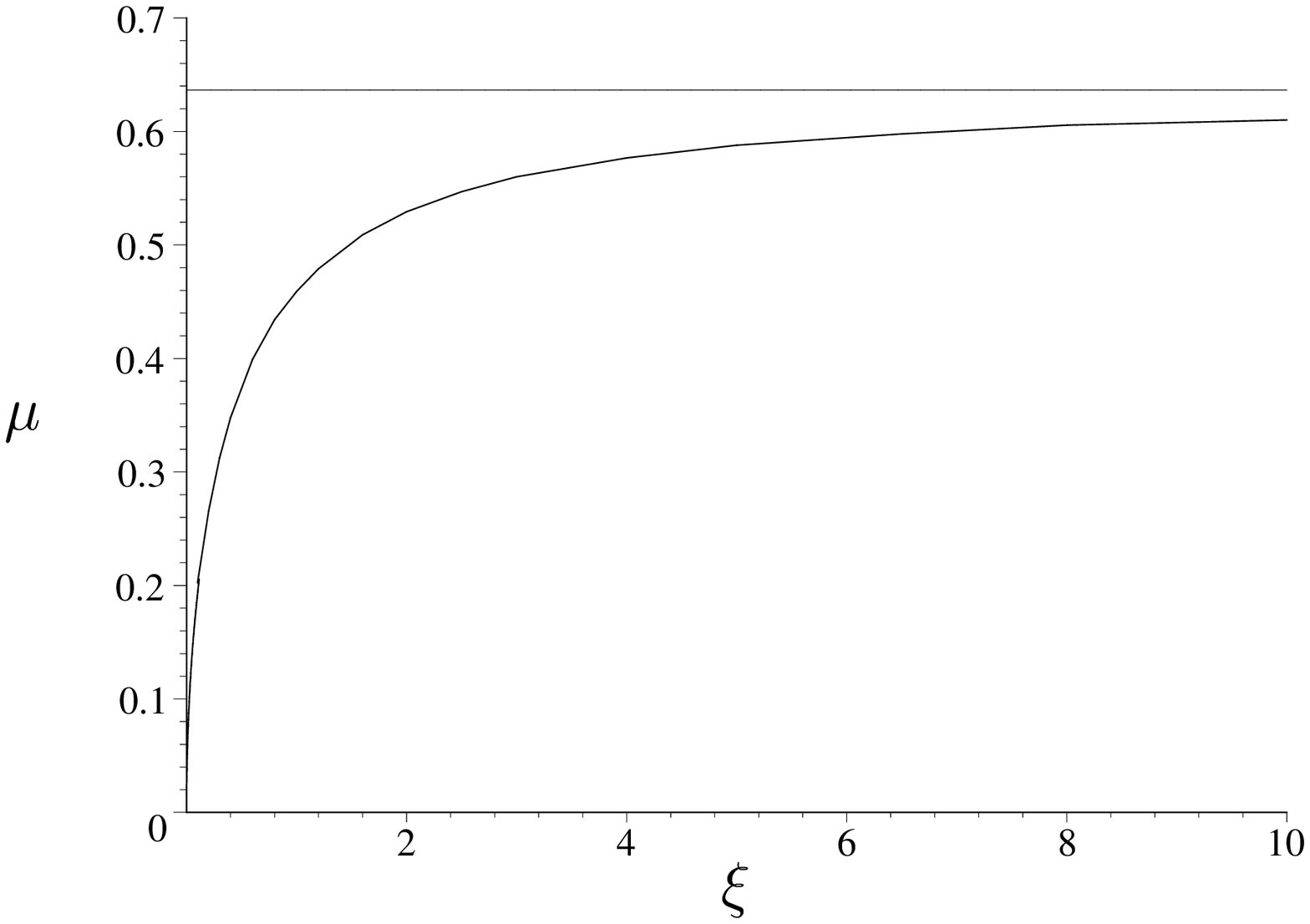,height=8cm,width=6.4cm,angle=270}
\caption{First order phase boundary separating the crystal from the massive Fermi gas phases at $T=0$ in the generalized GN model.
The vertical axis may be interpreted either as critical chemical potential or twice the baryon mass, due to fractional baryon number 1/2
in this model. The straight line shows the asymptotic value $2/\pi$ taken from the standard GN model. Numerical calculations performed
for a few extra points ($\xi=0.3,0.6,1.0,1.6,2.5,4.0,6.5,8.0$) in addition to the values mentioned in the caption of Fig.~\ref{fig8}.}
\label{fig9}
\end{center}
\end{figure}

\begin{figure}
\begin{center}
\epsfig{file=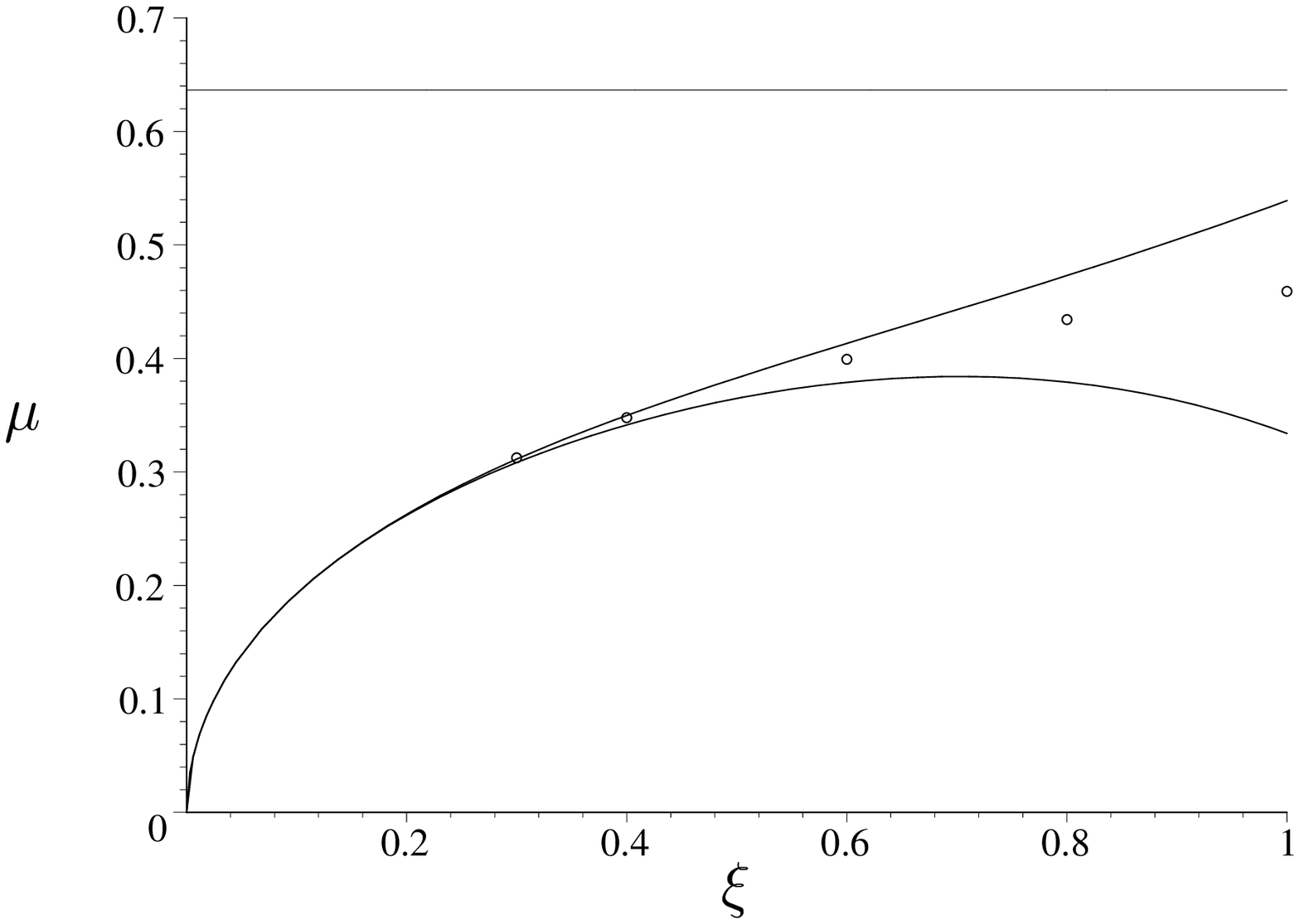,height=8cm,width=6.4cm,angle=270}
\caption{Like Fig.~\ref{fig9}, but blowing up the region of small $\xi$ to check the consistency between the derivative expansion of Sec.~\ref{sect4}
(lower curve: NNNLO, upper curve: NNLO) and the numerical HF calculation (circles).}
\label{fig10}
\end{center}
\end{figure}

\begin{figure}
\begin{center}
\epsfig{file=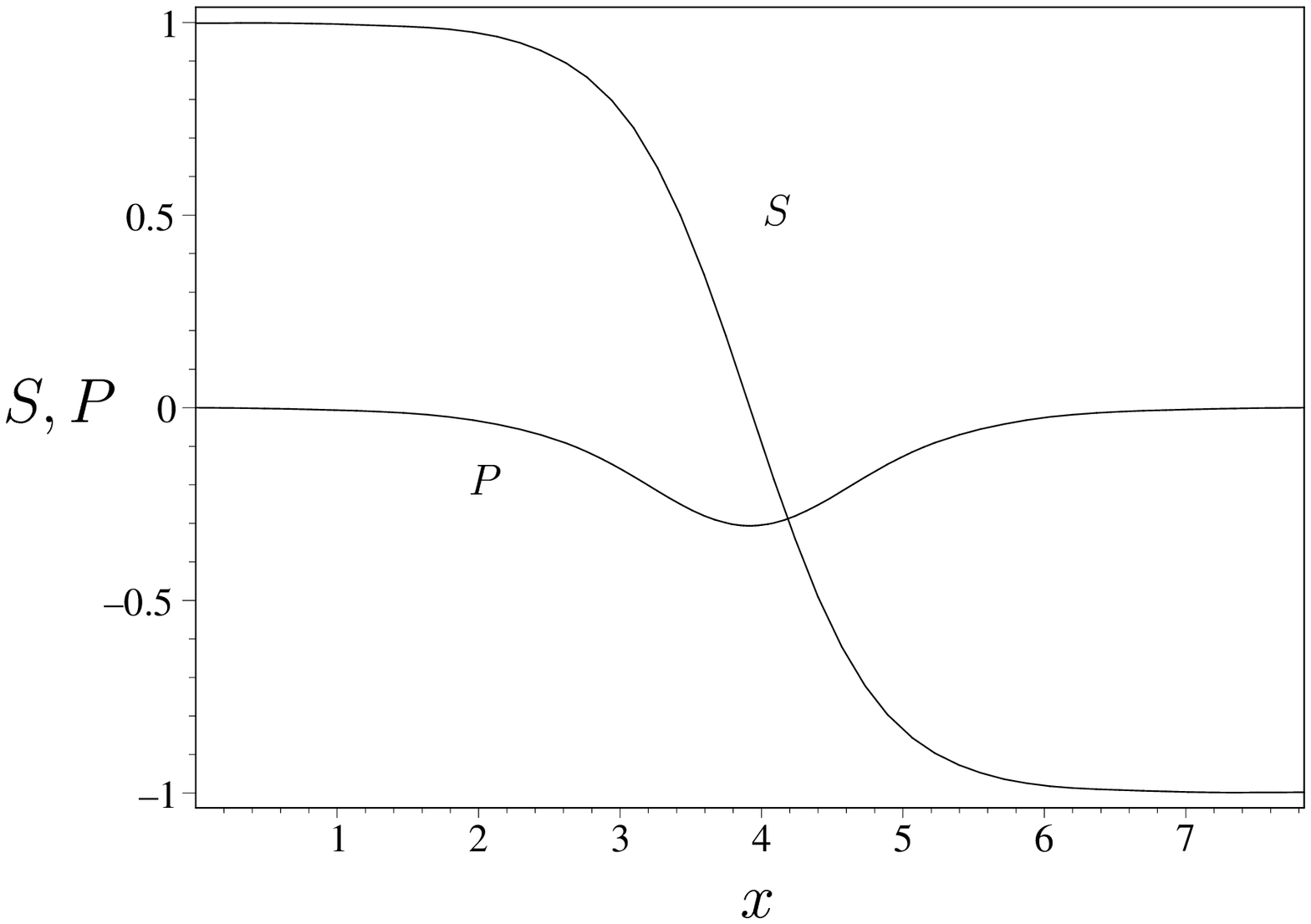,height=8cm,width=6.4cm,angle=270}
\caption{Example of numerical baryon HF potentials at $\xi=2.0$. For larger values of $\xi$, $P$ decreases and $S$ approaches the
GN kink (not shown).} 
\label{fig11}
\end{center}
\end{figure}

Let us first consider the 1st order critical line at $T=0$, i.e., the baseline of the 1st order critical sheet in a 3d plot. This is closely related
to the baryon mass discussed in Sec.~\ref{sect4} near the chiral limit. Since we are not restricted to small $\xi$ values in the numerical HF 
calculation, we can now get complementary information to the one of Sec.~\ref{sect4} and complete the picture about baryons in 
the generalized GN model. Fig.~\ref{fig9} shows the phase boundary at zero temperature in the ($\xi,\mu$) plane (the actual calculation
was done at $T=0.05$, but this makes no difference). Since baryon number is 1/2 in our model, the critical chemical potential has to be identified
with twice the baryon mass (divided by $N$) here. The reason is the following: The critical chemical potential at $T=0$ is the amount of
energy needed to add a fermion to the vacuum. If the kink-like baryon has mass $M_{\rm B}$ and carries $N/2$ fermions, we get 
$\mu_{\rm crit}=2 M_{\rm B}/N$. 
The curve in Fig.~\ref{fig9} interpolates between the massless baryons of the NJL$_2$ model and twice the mass 
of the kink in the GN model, $M_{\rm B}/N=1/\pi$. As shown in Fig.~\ref{fig10}, at small values of $\xi$ the numerical HF results match nicely
onto the derivative expansion, a welcome test of both the analytical and numerical approaches. 
From the HF calculation at the phase boundary we can also extract
the shape of the self-consistent potentials for a single baryon, now for arbitrary values of $\xi$. A typical example is shown in Fig.~\ref{fig11}
for the case $\xi=2$. The scalar potential has kink shape at all $\xi$, going over into the GN model kink in the limit $\xi \to \infty$. The pseudoscalar
potential is bell shaped and gets more and more suppressed with increasing $\xi$. This is of course just the effect of the double counting correction
term (\ref{4.1}) where $\xi$ acts like a Lagrange multiplier for $P$, quenching it completely in the limit $\xi\to \infty$. The other limit, $\xi \to 0$, has already 
been discussed before in Sec.~\ref{sect4} in terms of the sine-Gordon kink with scalar and pseudoscalar potentials of the same amplitude.

\begin{figure}
\begin{center}
\epsfig{file=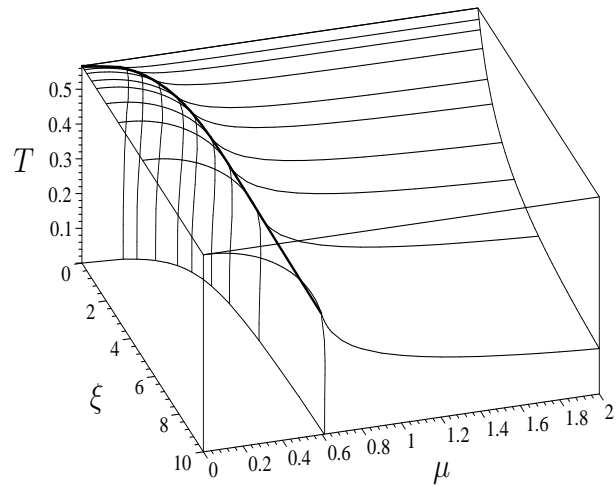,height=8cm,width=6.4cm,angle=270}
\caption{Like Fig.~\ref{fig8}, but including 1st order phase boundaries separating the crystal from the massive Fermi gas. The curve drawn at $\mu=2$ is
the asymptotic expectation according to Eq.~(\ref{7.14}), the base line at $T=0$ is taken from Fig.~\ref{fig9}.}
\label{fig12}
\end{center}
\end{figure}

\begin{figure}
\begin{center}
\epsfig{file=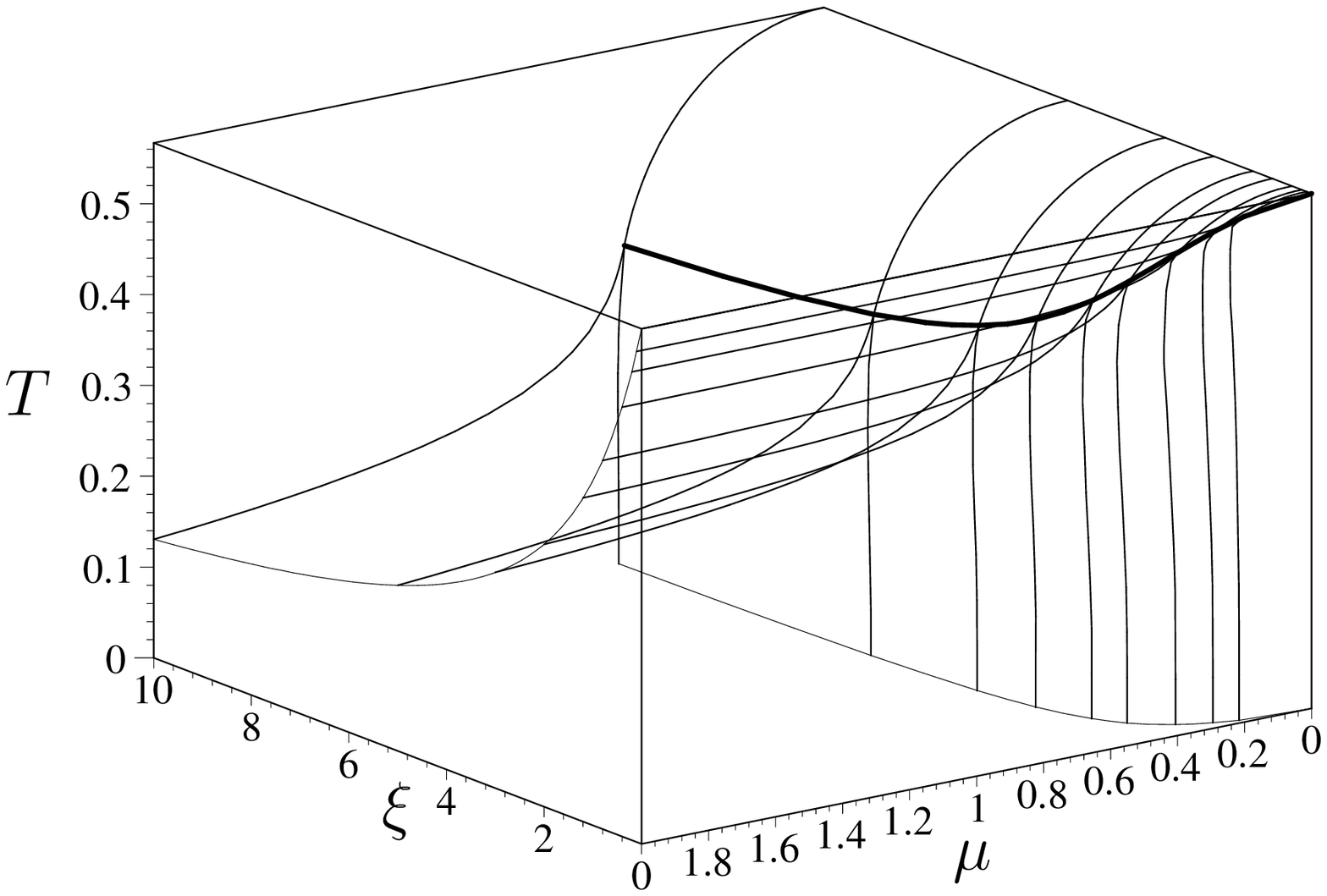,height=8cm,width=6.4cm,angle=270}
\caption{Like Fig.~\ref{fig12}, but different orientation for better visibility}
\label{fig13}
\end{center}
\end{figure}

Finally, we come to the full phase diagram as a function of $\xi,\mu,T$, including the numerically determined 1st order sheet.
It is shown in Fig.~\ref{fig12} and Fig.~\ref{fig13} under 2 different viewing angles for the sake of clarity. As explained in more detail in Ref.~\cite{8},
the phase boundary is determined by performing the HF calculation along a trajectory crossing the critical line and comparing the grand 
canonical potential of the massive Fermi gas to the one of the soliton crystal. As we know the exact location of the tricritical point in the present
case, we are even in a somewhat better position here than in the previous study of the massive NJL$_2$ model.

\section{Summary and conclusions}\label{sect8}
In this paper, we have studied a generalization of the GN model with two different (scalar and pseudoscalar) coupling constants. This equips us
with an ``interpolating field theory" between the well-studied massless GN and NJL$_2$ models in a way which always keeps the discrete Z$_2$ chiral
symmetry intact. The continuous chiral symmetry of the NJL$_2$ model is only recovered for equal coupling constants, so that we now break chiral symmetry 
(explicitly) in a quite different manner than via the usual fermion mass term. Our motivation was primarily to get further insights into the solitonic aspects of 4-fermion theories in 1+1
dimensions which have been investigated intensely in recent years. 

The first insight is the emergence of the dimensionless parameter $\xi$ during the process of regularization and renormalization, in addition to the
familiar fermion mass. The basic relations, Eqs.~(\ref{2.11}), which generalize the standard gap equation remove all divergences encountered in
subsequent applications, both in the treatment of bound states (mesons, baryons) and in the thermodynamics of the model. The parameter $\xi$ plays
a role analogous to the ``confinement  parameter" $\gamma$ in massive GN models. This is particularly striking in the RPA approach to the pseudoscalar 
fermion-antifermion bound and scattering states, where the results for the massive NJL$_2$ model and the generalized GN model become identical
if we replace $\gamma$ by $\xi$. The qualitative effect of $\xi$ on the HF calculations at zero and finite temperature is very easy to understand. 
It only enters in the double counting correction to energy or thermodynamic potential as an extra term $\sim \xi \int {\rm d}x  P^2$. Hence it acts like a
Lagrange multiplier for the pseudoscalar potential, leading to a complete quenching of $P$ in the GN limit $\xi\to \infty$. Thus $\xi$ may be thought of as
a ``chiral quenching parameter" responsible for the transition from complex condensates living on the chiral circle in the NJL$_2$ model to the purely real
condensates of the GN model.

As far as baryon structure is concerned, the most interesting result is perhaps the fact that the new baryons interpolate between the kink of the GN
model and the massless baryon of the NJL$_2$ model, always carrying fractional baryon number 1/2. This is certainly a consequence of the fact
that the generalized GN model still has a discrete chiral symmetry. Indeed in the massive NJL$_2$ model, chiral symmetry is explicitly broken 
by the mass term without a residual Z$_2$ symmetry and one finds baryons with integer baryon number 1. This new kind of chiral kink is different
from all known multi-fermion bound states in the GN model family and has been determined analytically for small $\xi$ and numerically for large $\xi$.

The phase diagrams of the NJL$_2$ and GN model look very different, so that we were curious to see how our theory would manage to interpolate 
between these two pictures. This can now be answered most clearly by the study of the tricritical behavior near the chiral limit, largely
analytically owing to the GL approach. The relevant picture is Fig.~\ref{fig5}, showing a kind of ``morphing" from GN-type behavior to the NJL$_2$
phase diagram with its single straight line phase boundary. Together with the numerical HF calculation, we are now confident
that the solitonic crystal phase is separated from the massless (massive) Fermi gas by a 2nd (1st) order transition, respectively. This was not clear
a priori, since the transition from the crystal to the massive homogeneous phase is continuous in the GN model and doesn't even exist in the NJL$_2$
model. Our interpolated phase diagram also looks qualitatively different from the one of the massive NJL$_2$ model which has only 2 phases
(no massless phase due to explicit breaking of the Z$_2$ symmetry), and where 
the opening angle between the 2 phase boundaries at the tricritical point was $\pi$ rather than 0. 

Initially, we had hoped that the generalized GN model can be solved analytically for arbitrary $\xi$, since this is what happens at the ``endpoints"
$\xi=0$ (NJL$_2$) and $\xi = \infty$ (GN). However, this does not seem to be the case. In this situation, the fact that
our toolbox also contains the numerical HF method has turned out to be a definite advantage. A combination of analytical
calculations and a numerical approach gives us confidence that we have solved and understood the model in the large $N$ limit fairly well. The
most serious limitation at present is the fact that our techniques are tailored to point-like 4-fermion interactions and cannot deal with gauge theories in a 
systematic fashion. This is unfortunate in view of the interesting features of, e.g., the 't~Hooft model \cite{21} where more analytical insights into
the early \cite{11} and very recent \cite{22,23} numerical HF calculations on the lattice would be welcome. 

\section*{Acknowledgement}

We should like to thank Gerald Dunne and Oliver Schnetz for stimulating discussions and their interest in this work.

\section*{Appendix: Details of the Ginzburg-Landau approach of Sec.~\ref{sect6}}

Here we collect the detailed formulae used in preparing Fig.~\ref{fig6} in Sec.~\ref{sect6}. We first list the coefficients of the Taylor expansion
(\ref{6.17}). 
Using the notation
\begin{equation}
\Psi^{(n)}(z_t) = R_n+ {\rm i} I_n, \qquad z_t = \frac{1}{2} + \frac{{\rm i}\tilde{\nu}_t}{2\pi},
\label{A1}
\end{equation}
one finds
\begin{eqnarray}
a_{20} & = & 0
\nonumber \\
a_{21} & = & 0
\nonumber \\
a_{22} & = & \frac{I_1^2-R_2}{4 \pi  T_t^2 \Omega^2} \sigma^2 + \frac{\Omega}{4\pi^2 T_t}\eta
\nonumber \\
a_{30} & = & - \frac{I_1}{8 \pi^2 T_t}
\nonumber \\
a_{31} & = & \frac{I_1^2-R_2}{8 \pi^2 T_t^2 \Omega}\sigma 
\nonumber \\
a_{32} & = & \frac{I_3 + 4 I_1 R_2 - 2 I_1^3}{16 \pi^2 T_t^3 \Omega^2} \sigma^2 + \frac{2\pi \zeta (I_1^2-R_2)+ R_2 \Omega^2 }
{16 \pi^3 T_t^2 I_1 \Omega}\eta
\nonumber \\
a_{40} & = & - \frac{R_2}{64 \pi^3 T_t^2}
\nonumber \\
a_{41} & = & \frac{2 I_1 R_2+I_3}{64 \pi^3 T_t^3 \Omega} \sigma
\nonumber \\
a_{42} & = & \frac{R_4 - 6 I_1 I_3 - 6 I_1^2 R_2}{128 \pi^3 T_t^4 \Omega^2} \sigma^2 
\nonumber \\
& & + \frac{2 \pi \zeta( I_3 + 2 I_1 R_2) -I_3 \Omega^2}{128 \pi^4 T_t^3 I_1 \Omega} \eta
\label{A2}
\end{eqnarray}
Eq.~(\ref{6.10}) now reads
\begin{equation}
\xi =  \frac{2\pi a_{30}^2}{a_{40}}=  - \frac{2 I_1^2}{R_2}.
\label{A3}
\end{equation}
The scale parameters $\lambda,\chi$ from Eq.~(\ref{6.26}) and the residual parameter $\kappa$ in the effective action (\ref{6.29})
then become,
\begin{eqnarray}
\lambda & = & \frac{8\pi (2 R_2^2+I_1I_3)}{\Omega R_2^2} \sigma
\nonumber \\
\chi^2 & = &  \frac{64 \pi^2 T_t I_1 (2 R_2^2+I_1I_3)}{\Omega R_2^3} \sigma
\nonumber \\
\kappa & = & \frac{R_2^3 (\pi \sigma^2(R_2-I_1^2)- \eta T_t \Omega^3)}{4 \pi \sigma^2 (2 R_2^2+I_1I_3)^2}
\label{A4}
\end{eqnarray}
2nd order phase boundary in local coordinates $\sigma, \eta$, see Eq.~(\ref{6.19}),
\begin{equation}
\eta = \frac{\pi}{T_t \Omega^3} \left(  (R_2-I_1^2)- \frac{(2 R_2^2+I_1I_3)^2}{ R_2^3}\right) \sigma^2.
\label{A5}
\end{equation}
1st order phase boundary,
\begin{equation}
\eta = \frac{\pi}{T_t \Omega^3} \left( (R_2-I_1^2) +(2 + \sqrt{6}) \frac{(2 R_2^2+I_1I_3)^2}{ R_2^3}\right) \sigma^2.
\label{A6}
\end{equation}
These critical lines can easily be rotated back to the original coordinates, see Fig.~\ref{fig6} for some results.

\end{document}